\documentstyle[aps,prc,floats,epsfig]{revtex}
\def\nuc#1#2{\relax\ifmmode{}^{#1}{\protect\text{#2}}\else${}^{#1}$#2\fi}
\begin{document} 
\twocolumn[\hsize\textwidth\columnwidth\hsize\csname
@twocolumnfalse\endcsname
\draft
\title{Cross sections for nuclide production in 1 GeV proton-irradiated 
$^{208}$Pb}
\author{Yu.~E.~Titarenko, O.~V.~Shvedov, V.~F.~Batyaev, E.~I.~Karpikhin,    
V.~M.~Zhivun, \\ A.~B.~Koldobsky, R.~D.~Mulambetov, D.~V.~Fischenko,   
S.~V.~Kvasova, A.~N.~Sosnin}
\address{Institute for Theoretical and Experimental Physics,
B.Cheremushkinskaya 25,\\  117259 Moscow, Russia} 
\author{S.~G.~Mashnik, R.~E.~Prael, A.~J.~Sierk}
\address{Los Alamos National Laboratory, Los Alamos, NM 87545, USA}
\author{T.~A.~Gabriel}
\address{Oak Ridge National Laboratory, Oak Ridge, TN 37831, USA}
\author{M.~Saito}
\address{Tokyo Institute of Technology 2-12-1, O-okayama, Meguro-ku,
Tokyo 152, Japan}
\author{H.~Yasuda}
\address{Japan Atomic Energy Research Institute, Tokai, Ibaraki, 319-1195, Japan}
\date{\today}
\maketitle

\begin{abstract}
114 cross sections for nuclide production in a 1.0 GeV proton-irradiated 
thin $^{208}$Pb target have been measured by the direct 
$\gamma$-spectrometry method using a high-resolution Ge detector.
The $\gamma$ spectra were processed by the GENIE-2000 code. The 
ITEP-developed SIGMA code was used together with the PCNUDAT nuclear 
decay database to identify the $\gamma$ lines and to determine the 
cross sections. The $^{27}$Al(p,x)$^{22}$Na 
reaction was used to monitor the proton flux.    
Results of a feasibility study of the auxiliary $^{27}$Al(p,x)$^{24}$Na 
and $^{27}$Al(p,x)$^{7}$Be monitor reactions in the 0.07--2.6 GeV 
proton-energy range 
are presented as well.
Most of the experimental data have been 
analyzed by  the LAHET (with ISABEL and Bertini options), 
CEM95, CEM2k, INUCL, CASCADE, 
CASCADE/INPE, and YIELDX codes that simulate hadron-nucleus 
interactions. 
\end{abstract}

\vspace*{0.2cm}

]

\section{introduction}

Major advances in Pb-Bi reactor technology, as manifested in
the Russian-made
Alpha-class submarines, has made Pb-Bi technology attractive for 
accelerator-driven systems (ADS) to be used at facilities 
designed to transmute nuclear waste \cite{1,2,3}.
These Pb-Bi reactors would operate in an environment of 
high-energy radiation, thus motivating studies of the 
nucleonics characteristics of lead and bismuth, including the 
yields of residual product nuclei under irradiation by protons of 
energies ranging from a few MeV to 2--3 GeV.
Results of such studies are extremely important when designing 
even demonstration versions of ADS facilities.
Yields of residual product nuclei define important characteristics
of these facilities such as radioactivity (both current and residual), 
deterioration of corrosion resistance, yields of gaseous products, 
neutron ``poisoning'', etc. That is why several scientific groups 
have been studying the yields of residual product nuclei from Pb and Bi 
(both natural and isotope-enriched) bombarded with protons
in the desired energy range \cite{4,5,6,7,8,9,9a}.

Most of these studies have been made using $\gamma$ spectrometry on 
proton-irradiated natural or 
monoisotopic thin Pb and Bi samples (for example, p$ + ^{208}_{\;\;82}$Pb).
This technique makes it possible to determine the independent and cumulative 
yields of more than a hundred reaction products \cite{4,5,6,7}.
An alternative technique of magnetically separating the reaction products 
after a heavy-ion beam interacts with a liquid H target,
using reverse kinematics, as done recently at GSI to study
$^{238}_{\;\;92}$U, 
$^{197}_{\;\;79}$Au and $^{208}_{\;\;82}$Pb$  + ^1_1$H, makes it 
possible to determine about a thousand yields \cite{8,9,9a}.

Comparing the results of the two techniques helps in understanding
the systematic errors inherent to both. This work makes use of the 
$\gamma$-spectrometry technique. We discuss below our  
results as well as the systematic errors of $\gamma$ spectrometry.

Necessarily, computational methods will play an important role 
in the designs for ADS facilities. Therefore, testing of the most 
extensively used simulation codes is necessary.
We perform a qualitative and quantitative comparison of our data with
simulated results from seven codes widely used in applications.

\section{Basic definitions and formulas}

The formalism for finding the reaction product yields (cross sections) 
in high-energy proton-irradiated thin targets is described in sufficient 
detail in \cite{6}. Any of the measured reaction products generated in 
nucleon interactions with matter are assumed to originate both
in the reaction proper and in the decays of its chain precursors.
Thus, the set of terms invented earlier in studying the mass and charge 
distributions of fission products 
(see, e. g., \cite{fiss}) 
can conveniently be used when processing 
and interpreting the experimental results of the present work. In this 
terminology, the independent and cumulative yields of products underlie
the formalism.

The independent yield of a reaction product nuclide is a cross section 
for the nuclide to be produced directly in the reaction, whereas the 
cumulative yield of the nuclide is a cross section for the nuclide to be 
produced in all the appropriate processes, i.e., both directly in the 
reaction and over the time during the decays of all of its chain 
precursors.\footnote
{Some applications also make use of the mass yield.
In its simplified form (disregarding alpha transitions and delayed-
neutron emission), the mass yield is the sum of all independent 
yields of the isobars of a given mass or equivalently is the sum 
of cumulative yields of the stable isotopes of a given mass.}

The variations in the concentration  of any two nuclides 
of a chain produced in an irradiated target 
$(N_1 \stackrel{\lambda_1}{\longrightarrow} N_2 
\stackrel{\lambda_2}{\longrightarrow})$
may be modeled as a set of differential equations that describe the 
production and decays of the nuclides.
By introducing the time functions, $F_i$, of the type  
$F_i=\left( 1-e^{-\lambda_i \tau} \right) 
\frac{\displaystyle 1-e^{- \lambda_i KT}}
{\displaystyle 1-e^{- \lambda_i T}}\mbox{ ,} \quad$ 
($i = 1, 2$, or monitor product  (Na or another)); $\tau$ is the duration 
of a single proton pulse, $T$ is the pulse repetition period,
$K$ is the number of irradiation pulses which characterize the 
nuclide decays within the irradiation time, and by expressing 
(similar to the relative measurements) the proton fluence via 
the monitor reaction cross section $\sigma_{st}$, we can present 
the cumulative and independent yields as
\begin{equation}
\label{1}
\sigma_1^{cum} = \frac{A_0}{\eta_1 \varepsilon_1 F_1 N_{Na}}
\frac{N_{Al}}{N_T} \frac{F_{Na}}{\lambda_{Na}}\; \sigma_{st} \mbox{ ,}
\end{equation}
or, if measurements of the second nuclide alone are used, as
\begin{eqnarray}
\sigma_1^{cum} &=& \frac{A_1}{\nu_1 \eta_2 \varepsilon_2 F_1 N_{Na}}
\frac{N_{Al}}{N_T} \frac{\lambda_2 - \lambda_1}{\lambda_2}
\frac{F_{Na}}{\lambda_{Na}}\; \sigma_{st} \mbox{ ,}
\label{1a} 
\\
\sigma_2^{ind} &=& \left( \frac{A_2}{F_2} + \frac{A_1}{F_1}
\frac{\lambda_1}{\lambda_2} \right)
\frac{1}{\eta_2\varepsilon_2 N_{Na}}
\frac{N_{Al}}{N_T} \frac{F_{Na}}{\lambda_{Na}}\; \sigma_{st} \mbox{ ,}
\label{2}  
\\
\sigma_2^{cum} &=& \sigma_2^{ind} + \nu_1 \sigma_1^{cum} = \nonumber \\
&=& 
\displaystyle 
\left( \frac{A_1}{F_1} + \frac{A_2}{F_2} \right)
\frac{1}{\eta_2 \varepsilon_2 N_{Na}}
\frac{N_{Al}}{N_T} \frac{F_{Na}}{\lambda_{Na}}
\; \sigma_{st}  \mbox{ .} 
\label{3} 
\end{eqnarray}
Here, $\sigma_1^{cum}$ is the 
cumulative \footnote{In case of absence of short-lived precursors, 
the yield of the nuclide $N_1$ should be considered as independent.}
cross section of the first 
nuclide $N_1$; $\sigma_2^{ind}$ and $\sigma_2^{cum}$  are the 
independent and cumulative cross sections of the second nuclide 
$N_2$; $N_{Al}$  and $N_T$  are the numbers of nuclei in the monitor 
and in the experimental target, respectively;
$N_{Na}$ is the number of Na nuclei produced in the monitor; 
$\eta_1$ and $\eta_2$ are $\gamma$-line yields; $\varepsilon_1$ and 
$\varepsilon_2$ are the spectrometer efficiencies at energies 
$E_{\gamma_1}$ and $E_{\gamma_2}$; $\nu_1$ is the branching 
ratio of the first nuclide; $\lambda_1$, $\lambda_2$, and $\lambda_{Na}$ are, 
respectively, the decay constants of the first and second 
nuclides and of the monitor product ($^{22}${Na} and/or $^{24}${Na}).

The factors $A_0$, $A_1$, and $A_2$ are calculated through fitting 
the measured counting rates in the total absorption peaks, which 
correspond to energies $E_{\gamma_1}$ (the first nuclide) and 
$E_{\gamma_2}$ (the second nuclide), by the exponential functions
\begin{equation}
\label{exp}
g(t) = A_0 \frac{1-e^{- \lambda_1 t_{true}}}{\lambda_1 t_{true}} e^{- \lambda_1 t} 
\end{equation}
and
\begin{eqnarray}
\label{exp2}
\displaystyle
f(t) &=& A_1 \frac{1-e^{- \lambda_1 t_{true}}}{\lambda_1 t_{true}} e^{- \lambda_1 t}
 + \nonumber  \\
 &+&  \; \displaystyle
 A_2  \frac{1-e^{- \lambda_2 t_{true}}}{\lambda_2 t_{true}} e^{- \lambda_2 t}  \mbox{ .}		
\end{eqnarray}

Here, $t_{true}$ is the real time over which the spectrum is
measured, $t$ is the time between
the end of the irradiation and the beginning of the spectrum measurement.
 The forms of functions 
(\ref{exp}) and (\ref{exp2}) 
are the similar to what is presented in \cite{6}, but we present them here for clarity:
\begin{eqnarray}
A_0 &=& N_T \Phi \sigma_1^{cum} \eta _1 \varepsilon_1 F_1 \mbox{ ,}		
\label{expA0}
\\	
A_1 &=& N_T \Phi \sigma_1^{cum} \eta_2 \varepsilon_2 F_1 \nu_1
\lambda_2 / ( \lambda_2 - \lambda_1 ) \mbox{ ,}
\label{expA1}
\\		
A_2 &=& N_T \Phi \eta_2 F_2 [ \sigma_2^{ind}
- \sigma_1^{cum} \nu_1 \lambda_1 / ( \lambda_2 - \lambda_1 )]	\mbox{ .}	
\label{expA2}
\end{eqnarray}

\begin{figure}[t]
\centerline{\epsfxsize 9cm \epsffile{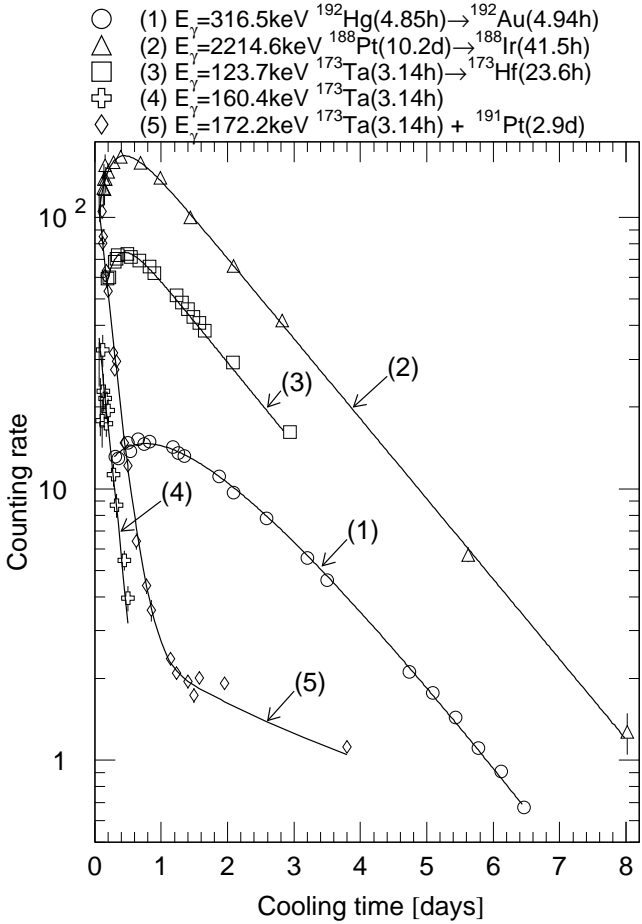}}
\vspace*{2mm}
\caption{
Typical examples of decay curves. Curve (1) is for the chain 
$^{192}$Hg $\to$ $^{192}$Au. Curve (2) is for 
$^{188}$Pt $\to$ $^{188}$Ir.
Curve (3) is for $^{173}$Ta $\to$ $^{173}$Hf. Curve (4) is 
for the independent $^{173}$Ta decay. Curve (5) is for the 
independent ($^{173}$Ta + $^{191}$Pt) decay.
Curve (1) is drawn with a time scale factor of 
4 and a counting rate factor of 0.1.
Curve (2) is drawn with a time scale factor of 0.1
and a counting rate factor of 100.
}
\label{fig1}
\end{figure}

It should be noted that formulas (\ref{1})--(\ref{3}) 
are deduced on the assumption that the corresponding $\gamma$ counting 
rates of each nuclide produced under irradiation are determined to within 
the desired accuracy throughout the time interval from the 
irradiation end to 
the ultimate intensity detection threshold. Curve 1 of 
Fig.~\ref{fig1} exemplifies such a favorable situation.
The situation corresponds to the $^{192}$Hg (T$_{1/2}$ = 4.85 hours)
$\to$ $^{192}$Au (T$_{1/2}$ = 4.94 hours) decay 
chain, with the $^{192}$Au, 316.5 keV $\gamma$ line being measured.
Despite the very similar periods, the absence of any addends to the 
measured $\gamma$ line (i.e., null contribution from     
$\gamma$ lines of any other nuclides whose energy is the same, within 
the spectrometer resolution, as of the measured $\gamma$ line) 
provides a sufficiently accurate determination of the cumulative 
$^{192}$Hg yield, as  well as the independent and cumulative 
$^{192}$Au yields.

The real situation often gets more complicated, however, because the 
decay curve of the second nuclide cannot be measured correctly in every 
case within the desired (in the sense as mentioned above) time interval.
Its $\gamma$-line intensity is often difficult to measure in the beginning 
of a measurement run, right after the irradiation stops. In the case of 
short-lived nuclides, this is due to the fact that within the period from 
the end of irradiation to the beginning of a measurement run (the cooling 
time), the nuclide $N_1$ (precursor) can partly or fully decay (if 
$\lambda_1 > \lambda_2$), or else a full or partial equilibrium sets in 
(if $\lambda_1 < \lambda_2$) between the nuclides $N_1$ and $N_2$.

The contribution from nuclide $N_1$ will then never be reflected in the 
experimental decay curve of nuclide $N_2$. The situation gets even more 
complicated because a great number of the products 
can include 
nuclides whose half-lives are close to the half-life of a shorter-lived 
nuclide, either $N_1$ or $N_2$. In that case, as noted above, if the 
$\gamma$-line energy of such nuclides is the same (within the spectrometer 
resolution) as the measured $\gamma$-line energy of 
other nuclide,
then 
neither of the factors $A_1$ and $A_2$ can be determined, or else the two 
can only be determined with great uncertainty and thus become useless for 
calculating the yields.

Fig.~\ref{fig1} is also a good illustration of a possible unfavorable 
situation for analyzing the nuclides of similar half-lives (Curve 1).
If the $^{192}${Au} decay curve begins being measured more than two days 
after the irradiation, the $^{192}${Hg} contribution becomes uncertain, 
resulting in erroneous calculations of the $^{192}${Au} yield. Such a 
situation is most probably responsible for the difference in the measured 
$^{192}${Au} yields of more than a factor 3 between the data of the present 
work (see Table II below) and the results of \cite{4} 
(46.9$\pm$6.6 mb and 160$\pm$50 mb, respectively).

Analyzing possible structures of radioactive chains permits the following 
two very common situations to be singled out.

First, assume that $\lambda_1<\lambda_2$. This situation is exemplified in 
Fig.~\ref{fig1} (Curve 2), which shows the decay curve of chain nuclides 
$^{188}${Pt}(T$_{1/2}$=10.2 days)$\to$$^{188}${Ir}(T$_{1/2}$=41.5 hours) 
recorded by measuring the 2214.6 keV $\gamma$ line of the $^{188}${Ir} 
daughter nuclide. In this case, the decay curve of nuclide $^{188}${Ir} 
can be used to obtain fairly accurate values of the factors $A_1$ and $A_2$ 
and, hence, of $\sigma_1^{cum}$, $\sigma_2^{ind}$, and $\sigma_2^{cum}$ 
($\sigma_{^{188}{Pt}}^{cum}$, 
$\sigma_{^{188}{Ir}}^{ind}$, and $\sigma_{^{188}{Ir}}^{cum}$).
Should this favorable situation get complicated (for example, 
the measurements began in 2 days after irradiation), then, even without 
observing the knee, the conclusion concerning the $^{188}${Pt} production 
is quite obvious because the 2214.6 keV $\gamma$ line of $^{188}${Ir}  is 
measured with the $^{188}${Pt} period. In this case, formula (\ref{1a}) may 
be used to calculate the $\sigma_{^{188}{Pt}}^{cum}$ value, whereas the 
$^{188}${Ir} data prove to be lost.

It should be noted that there even exists a nuclear data library that 
presents the $^{188}${Pt} gamma yields corrected for the 
$(\lambda_2-\lambda_1)/\lambda_2$ value. The corrected yields are marked 
with an index (D) to notify the validity of using the daughter nuclide 
$\gamma$ lines when calculating the number of mother nuclei \cite{10}.

The inverse situation ($\lambda_1 > \lambda_2$) permits the factor $A_2$ 
alone to be determined reliably. This is exemplified by Curve 3 in 
Fig.~\ref{fig1}, which is the decay curve of the chain nuclides 
$^{173}${Ta} (T$_{1/2}$=3.14 hours)$\to$$^{173}${Hf} (T$_{1/2}$=23.6 hours) 
recorded by measuring the 123.7 keV $\gamma$ line of $^{173}${Hf}. In this 
case, the factors $A_1$ and $A_2$ can be determined to within the desired 
accuracy, and formulas (\ref{1a})--(\ref{3}) can be used to calculate 
$\sigma_1^{cum}$, $\sigma_2^{ind}$, and $\sigma_2^{cum}$ 
($\sigma_{^{173}{Ta}}^{cum}$, $\sigma_{^{173}{Hf}}^{ind}$, and 
$\sigma_{^{173}{Hf}}^{cum}$). In cases when unfavorable situations arise 
due to, for example, measurements starting a day after the irradiation, the 
factor A$_1$ alone can be determined.

If, however, the eigen $\gamma$ line is used (this is shown in 
Fig.~\ref{fig1} by Curve 4, which is the $^{173}${Ta} decay curve measured 
by the 160.4 keV $^{173}${Ta} $\gamma$ line, or by Curve 5, which is the 
same, but inferred from the 172.3 keV $^{173}${Ta} $\gamma$ line with an 
addend contributed by the $^{191}${Pt} decay), then the factor $A_0$ can 
be calculated and the missing factor A$_1$ is
\begin{equation}
\label{expA11}
A_1 = A_0 \frac{\eta_2 \varepsilon_2}{\eta_1 
\varepsilon_1} \nu_1 \frac{\lambda_2}{\lambda_2-\lambda_1} \mbox{ ,}
\end{equation}
whereupon formulas (\ref{1a})--(\ref{3}) can be used to find 
$\sigma_1^{cum}$, $\sigma_2^{ind}$, and $\sigma_2^{cum}$ 
($\sigma_{^{173}{Ta}}^{cum}$, $\sigma_{^{173}{Hf}}^{ind}$, 
and $\sigma_{^{173}{Hf}}^{cum}$).
If, however, the factor $A_1$ cannot be found, we may use the factor 
$A_2$ together with expression (\ref{sigma2}) from \cite{6} to determine 
the quantity $\sigma_2^{cum^*}$, which we call the supra cumulative 
yield:
\begin{eqnarray}
\label{sig2} 
\displaystyle 
\sigma_2^{cum^*}&=&\sigma_2^{ind} + \frac{\lambda_1}{\lambda_1-\lambda_2} \nu_1 
\sigma_1^{cum} = \nonumber \\  
\displaystyle 
&=& \frac{A_2}{\eta_2\varepsilon_2F_2N_{Na}}
\frac{N_{Al}}{N_T}\frac{F_{Na}}{\lambda_{Na}}\sigma_{st} \mbox{ .}
\end{eqnarray}

It should be noted that the difference between $\sigma_2^{cum}$ and 
$\sigma_2^{cum^*}$ is not specified in many of the relevant publications 
despite the fact that $\sigma_2^{cum^*}$ is always greater than 
$\sigma_2^{cum}$. The explanation is that the formal equality
$\sigma_2^{cum}=\sigma_2^{cum^*}$
holds in case the generation times of the first and second nuclides are 
different. In this case, the time correction can be determined:
\begin{equation}
\label{delta}
\Delta t = \frac{1}{\lambda_2} \left[ 1+\frac{\nu_1 
\sigma_1^{cum} }{\sigma_2^{ind}+ \nu_1 \sigma_1^{cum} } 
\left( \frac{\lambda_2}{\lambda_1-\lambda_2}\right) \right] \mbox{ .}
\end{equation}

From formula (\ref{delta}), it is seen that $\Delta$t depends on the yields 
$\sigma_1^{cum}$ and $\sigma_2^{ind}$, thereby preventing the time 
correction $\Delta t$ from being allowed for in a general case when 
determining $\sigma_2^{cum}$. In common cases of 
$\sigma_2^{ind} \ll \nu_1 \sigma_1^{cum}$, however, the time correction 
$\Delta t$ can be shown to depend on the decay constants only:
\begin{equation}
\label{delta2}
\Delta t \cong \Delta t^{'} = \frac{1}{\lambda_2} \left[1+
\left(\frac{\lambda_2}{\lambda_1-\lambda_2}\right) \right] \mbox{ .}
\end{equation}

Therefore, the cumulative yield of the second nuclide can be determined 
after a post-irradiation period sufficient for $N_1$ to decay into $N_2$ 
(normally, this equals from 6 to 10 half-lives of the first nuclide) 
by measuring the decay curve of the first nuclide making allowance 
for the time correction $\Delta t$:
\begin{eqnarray}
\label{sigma2} 
\displaystyle 
\sigma_2^{cum} &=& \frac{A_2}{\eta_2 \varepsilon_2 F_2 N_{Na}}
\frac{N_{Al}}{N_T}\frac{F_{Na}}{\lambda_{Na}} \sigma_{st} 
e^{-\lambda_2 \Delta t}  \cong  \nonumber \\ 
\displaystyle
&\cong&\frac{A_2}{\eta_2 \varepsilon_2 F_2 N_{Na}}
\frac{N_{Al}}{N_T}\frac{F_{Na}}{\lambda_{Na}} \sigma_{st} 
\left(1-\frac{\lambda_2}{\lambda_1}\right) .
\end{eqnarray}

In the case where 
the condition $\sigma_2^{ind} \ll \sigma_1^{cum} \nu_1$ 
is not satisfied, so the $\sigma_2^{cum}$ value cannot be calculated in any 
way accurately, we may estimate the difference 
\begin{equation}
\label{sigma21}
\Delta \sigma_2^{cum^*}= \sigma_2^{cum^*} - \sigma_2^{cum} = 
\frac{\lambda_2}{\lambda_1-\lambda_2} \nu_1 \sigma_1^{cum} \mbox{ .}
\end{equation}

Proceeding from the condition 
$\sigma_2^{cum^*} \ge \sigma_2^{cum} \ge \sigma_1^{cum} \nu_1$, 
we may estimate the upper limit of $\Delta \sigma_2^{cum^*}$:
\begin{equation}
\label{sigma22}
\Delta \sigma_2^{cum^*} \le \frac{\lambda_2}{\lambda_1-\lambda_2} 
\sigma_2^{cum^*}
\end{equation}
or, in the relative form, 
\begin{equation}
\label{delt}
\delta \sigma_2^{cum} = \frac{\Delta \sigma_2^{cum^*}}
{\sigma_2^{cum^*}} \cdot 100\% \le \frac{\lambda_2}{\lambda_1-\lambda_2} 
\cdot 100\% \mbox{ .}
\end{equation} 

From the resultant formula (\ref{sigma22}) it is seen that the measured 
value of the supra cumulative yield $\sigma_2^{cum^*}$ may sometimes prove 
to be very different from its true value $\sigma_2^{cum}$.
In the case of $^{179}${Re}, for example, the data of the present work 
give a $\delta \sigma_2^{cum}$ value of $\sim 55\%$. This fact should 
be borne in mind when comparing between experimental and simulated data.

\section{Experimental techniques}

The used experimental techniques 
are described in \cite{6}.
A 10.5 mm diameter, 139.4 mg/cm$^2$ monoisotopic $^{208}${Pb} metal 
foil sample (97.2\% $^{208}${Pb}, 
1.93\% $^{207}${Pb}, 0.87\% $^{206}${Pb}, $<$0.01\% $^{204}${Pb}, 
$<$ 0.00105\% of chemical impurities) was irradiated by protons.
A 139.6 mg/cm$^2$ Al foil of the same diameter was used as monitor. Chemical 
impurities in the monitor do not exceed 0.001\%.

The sample was irradiated by the 995 MeV external proton beam of the ITEP 
U-10 synchrotron described  briefly in \cite{6}. The irradiation time was
an hour. The total proton fluence was $5\cdot10^{13}$p/cm$^2$. The 
$\gamma$-spectrometer measurements began 10 minutes after the irradiation 
stopped. The spectra have been measured for half a year.

It should be noted that the experimental program of measuring the reaction 
product yields from diverse materials was supported by other studies aimed 
at reducing the systematic errors in the experimental results 
\cite{27}. These studies include:
\begin{itemize}
\item 
experiments to specify the neutron component in the extracted proton beams,
\item 
experiments to specify the monitor-reaction cross section,
\item 
experiments to specify the variations of the $\gamma$-spectrometer detection 
efficiency, depending on the height of the irradiated sample position above 
the detector,
\item 
studies to optimize the $\gamma$-spectrum processing codes.
\end{itemize}

The results of these studies were used in determining the cross sections 
for nuclide production in the measured $^{208}${Pb}(p,x) reactions.

\subsection{Neutron component in the extracted proton beams}

The proton beams extracted from accelerators include not only the primary 
protons, but also secondary particles 
(neutrons, protons, $\pi$ mesons, and gammas) 
produced in the primary proton interactions with the structural materials of 
transport channels and shielding.

Identical reaction products can be produced in interactions of 
various secondaries with an experimental sample. Since particular 
nuclear reactions that generate a given nuclide cannot be identified in 
the 
measurements, the extracted proton beams have to be tested 
and specified thoroughly.

Track detectors were first used in the experiments to discriminate the 
neutron component in the proton beams \cite{6}. Later, direct 
$\gamma$ spectrometry was used for the purpose.
Solid-state nuclear track detectors (SSNTD) of an improved geometry with 
a collimating grid and a glass fragment track detection material were used 
to record the fission fragments from a fissile layer, thereby improving the 
absolute detector efficiency.

An SSNTD with a 61.5 $\mu$g/cm$^2$ $^{209}${Bi} layer was used to measure 
the proton flux density. 
$^{209}$Bi was selected because 
the cross 
section for its fission induced by secondary neutrons is small compared 
with that for primary protons
($\overline{\sigma_{^{209}Bi(n,f)}}<<\sigma_{^{209}Bi(p,f)}$).
The neutron flux density was measured using an SSNTD with a 880 
$\mu$g/cm$^2$ $^{237}${Np} layer.

The following experimental design was adopted. The extracted proton beam 
irradiates a $^{209}${Bi}-containing ``sandwich" 
(Bi layer + collimator + glass), while similar sandwiches with $^{237}${Np} 
layers are placed along a line normal to the beam axis 
at distances of 20--435 mm from the axis.

In the experiments, the neutron-to-proton flux density ratio, 
$\Phi_n/\Phi_p$, was measured 
using the expression
\begin{equation}
\label{neutr}
\frac{\Phi_n}{\Phi_p} = \frac{T_1}{T_2} \cdot \frac{\sigma^{^{209}Bi}_{p,f}}
{\overline{\sigma}^{^{\;\;237}Np}_{n,f}} \cdot \frac{N^{^{209}Bi}}
{N^{^{237}Np}} \cdot \frac{\xi_2}{\xi_1} \mbox{ ,}
\end{equation} 
where $T_1$ and $T_2$ are numbers of measured tracks of $^{237}${Np} 
and $^{209}${Bi} fission products, respectively; 
$N^{^{237}Np}$ 
and 
$N^{^{209}Bi}$ 
are numbers of the $^{237}${Np} and $^{209}${Bi} nuclei, 
respectively; $\xi_1$ and $\xi_2$ are, respectively, 
corrections to the $^{237}${Np} and  $^{209}${Bi} layers, which 
allow for the anisotropy of fission-fragment 
ejection and for the variations of the solid angle of fission-fragment 
ejection through the collimator grid; $\sigma^{^{209}Bi}_{p,f}$ is the 
cross section for proton-induced $^{209}${Bi} fission;  
$\overline{\sigma}^{^{\;\;237}Np}_{n,f}$ is the weighted mean 
$^{237}${Np} neutron-induced fission cross section calculated as
\begin{equation}
\label{weight}
\overline{\sigma}_x = \frac{\int \sigma_x(E) \Phi(E) dE}{ \int \Phi(E) 
dE} \mbox{ ,}
\end{equation} 
where $x$ stands for 
$^{237}${Np}(n,f), $^{27}${Al}(n,p)$^{27}${Mg}, $^{27}${Al}(n,x)$^{24}${Na}, 
and $^{27}${Al}(n,x)$^{22}${Na}.

The fission cross section $\sigma^{^{\;\;237}Np}_{n,f}(E)$
needed to calculate $\overline{\sigma}^{^{\;\;237}Np}_{n,f}$
was retrieved from the WIND data library \cite{wind}.
The cross section for the proton-induced $^{209}${Bi} fission, 
$\sigma^{^{209}Bi}_{p,f}$, was taken from \cite{11}. 

\begin{figure}[t]
\vspace{1cm}
\centerline{\epsfxsize 8.0cm \epsffile{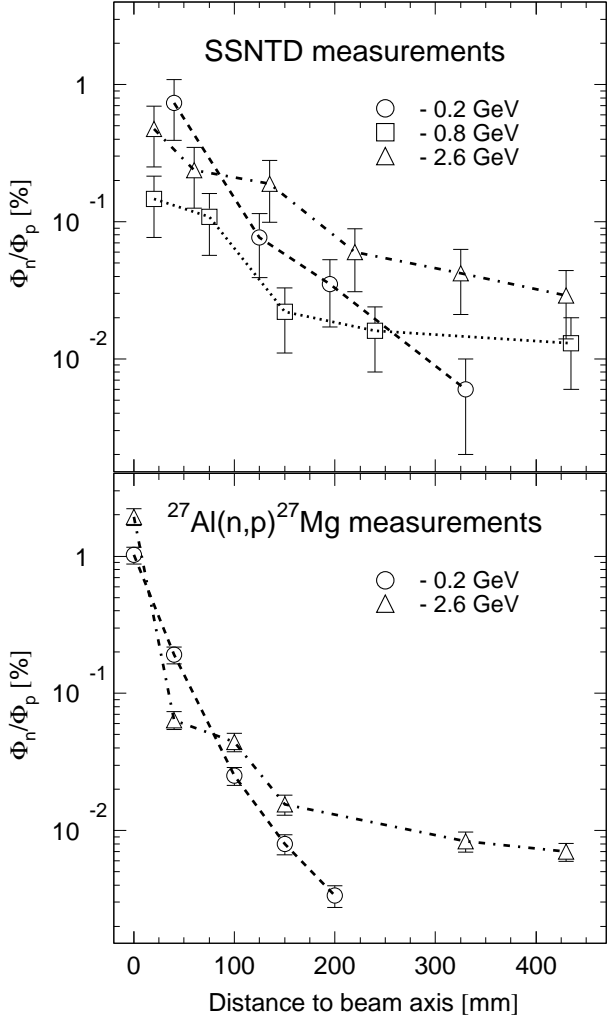}} 
\vspace*{2mm}
\caption{The neutron-to-proton mean flux density ratios 
versus distance from the proton beam axis calculated 
using: (1) the SSNTD measurements (upper plot);
(2) \nuc{27}{Al}(n,p)\nuc{27}{Mg} measurements (bottom plot).}
\label{fig2}
\end{figure}

The experiments were 
made with 200, 800, and 2600 MeV proton beams. Fig.~\ref{fig2} shows 
the resultant $\Phi_n/\Phi_p$ ratios as  functions of the perpendicular 
distance to the proton beam. The  $\Phi_n/\Phi_p$  ratio right in the 
proton beam was estimated by extrapolating the peripheral results to the 
center and proved to be of about (0.2--2)\%.

The feasibility of distinguishing the (n,p) reactions from (p,x) 
reactions has permitted an alternative pattern of direct 
$\gamma$ spectrometry. Namely, Al samples were irradiated, and 
$^{27}${Al}(n,p)$^{27}${Mg} (a $\sim$ 2.5MeV threshold), 
[$^{27}${Al}(n,$\alpha$)$^{24}${Na} (a $\sim$5.5 MeV threshold) + 
$^{27}${Al}(p,x)$^{24}${Na} (a $\sim$25 MeV threshold)], 
$^{27}${Al}($\displaystyle ^n_p$,x)$^{22}${Na}, 
and
$^{27}${Al}($^n_p$,x)$^{7}${Be} 
reaction rates were measured, in the beam 
center and at periphery.
The $^{27}${Al}(n,p)$^{27}${Mg} reaction characteristics have made it 
possible to detect $^{27}${Mg} in the experimental $^{27}${Al} samples 
positioned normally to the proton beam axis at the beam center and 
at distances of 40--430 
mm from the beam axis.
The \nuc{27}{Al}(n,p)\nuc{27}{Mg} cross sections were derived from 
the MENDL2 library \cite{12}.
The distance-dependent $\Phi_n/\Phi_p$ ratios measured via 
$^{27}${Al}(n,p)$^{27}${Mg} reaction are shown in Fig.~\ref{fig2} 
together with the results of the SSNTD method.
 
The technique has also made it possible to detect $^{27}${Mg} 
together with $^{24}${Na}, $^{22}${Na}, and $^{7}${Be}, when $^{27}${Al} 
was irradiated in the proton beam.
In this case, the neutron-to-proton flux density ratio in the beam is 
estimated as
\begin{equation}
\label{relat1} 
\frac{\Phi_n}{\Phi_p}=  \frac{\frac{
\displaystyle
\sigma^{^7Be,^{22}Na,^{24}Na}_{p,x}}
{
\displaystyle 
\overline{\sigma}^{^{\;\;27}Mg}_{n,p}}}
{\frac{
\displaystyle 
N^{^7Be,^{22}Na,^{24}Na}}{\displaystyle N^{^{27}Mg}}-
\frac{
\displaystyle 
\overline{\sigma}^{\;\;^7Be,^{22}Na,^{24}Na}_{n,x}}
{
\displaystyle 
\overline{\sigma}^{^{\;\;27}Mg}_{n,p}}} \mbox{ ,}
\end{equation} 
where $\overline{\sigma}^{^{\;\;27}Mg}_{n,p}, 
\overline{\sigma}^{^{\;\;22}Na}_{n,x}, 
\overline{\sigma}^{^{\;\;7}Be}_{n,x}$, 
and 
$\overline{\sigma}^{^{\;\;24}Na}_{n,\alpha}$ 
are the neutron spectrum-weighted cross sections of the above 
reactions calculated by formula (\ref{weight})
using the excitation functions from MENDL2 database \cite{12};
$\sigma^{^{22}Na}_{p,x}, 
\sigma^{^{24}Na}_{p,x}$, and $\sigma^{^{7}Be}_{p,x}$ 
are the $^{27}${Al}(p,x)$^{24}${Na}, $^{27}${Al}(p,x)$^{22}${Na}, and
$^{27}${Al}(p,x)$^{7}${Be} reaction cross sections; 
N$^{^{24}Na}$, N$^{^{22}Na}$, N$^{^{27}Mg}$, 
and N$^{^{7}{Be}}$ are numbers of the $^{24}${Na}, 
$^{22}${Na}, $^{27}${Mg}, and $^{7}${Be} nuclei produced in the Al samples 
with allowance for their decays under irradiation.

The techniques described above were used in the experiments with proton 
beams of 0.07, 0.1, 0.13, 0.2, 0.8, 1.0, 1.6, and 2.6 GeV.
 
\begin{figure}[t]
\vspace{1cm}
\centerline{\epsfxsize 8.0cm \epsffile{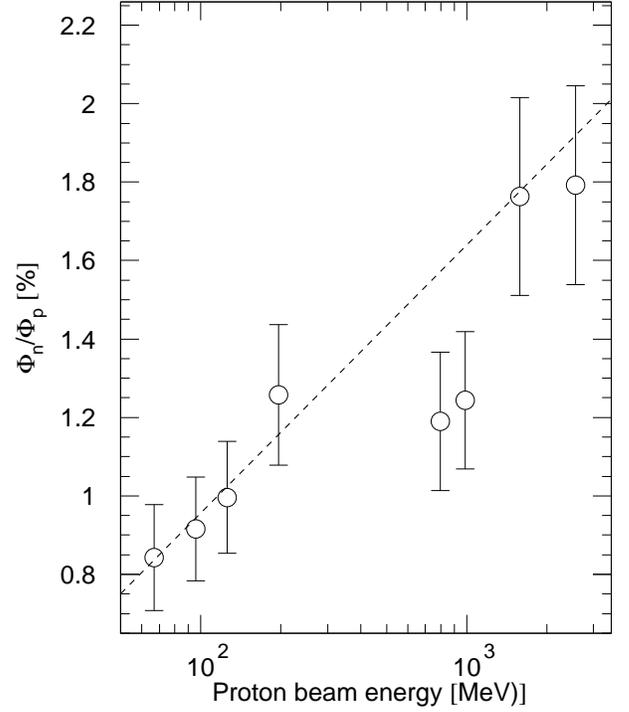}} 
\vspace*{1mm}
\caption{The neutron-to-proton mean flux density ratios in 
the proton beam versus proton energy as calculated 
from the $\gamma$-spectrometry data.}
\label{fig3}
\end{figure}

As seen from Figs.~\ref{fig2} and \ref{fig3}, 
the neutron component estimates obtained 
by the both techniques (SSNTD and $\gamma$ spectrometry) 
are essentially the same and are small.
It should be noted that the $^{208}${Pb}(p,x) reaction products measured 
at 1 GeV proton energy (see Table II) include none of the
nuclides 
producible in the (n,xn) reactions up to x = 5 ($^{204m}${Pb}), for such 
nuclides are either stable or long-lived. Our estimates have shown that, 
in the case of Pb isotopes of masses less than or equal to $^{204}${Pb}, 
additional production of daughter nuclei by (n,xn) reactions is 
below the level of experimental errors.

\subsection{Monitor reactions}

Contrary~to~\cite{6},~the~present~work~uses~the~$^{27}${Al}(p,x)$^{22}${Na} 
monitor reaction. Current practice suggests that three monitor 
reactions should be used on $^{27}${Al}, namely, 
$^{27}${Al}(p,x)$^{24}${Na} for short-term irradiations, 
$^{27}${Al}(p,x)$^{22}${Na} and $^{27}${Al}(p,x)$^{7}${Be}
for long-term irradiations.

Therefore, we have made additional experiments to study the $^{24}${Na} and 
$^{7}${Be} yields relative to $^{22}${Na}.
The $^{22}${Na} cross section values were obtained from \cite{tobailem}.
Since $^{24}${Na}, $^{7}${Be}, and $^{22}${Na} are produced in one and 
the same irradiated sample, their cross section ratios (represented by 
$\sigma^{^{24}{Na}}/\sigma^{^{22}{Na}}$, as an example) in a general 
form allowing for the neutron component can be calculated as
\begin{eqnarray}
\label{mon1} 
\frac{\sigma^{^{24}{Na},^{7}{Be}}}{\sigma^{^{22}{Na}}}&=&
\frac{A_0^{^{24}Na,^{7}{Be}}}{A_0^{^{22}Na}}
\frac{\left(\eta \varepsilon \right)^{^{22}Na}}{\left(
\eta \varepsilon \right)^{^{24}Na,^{7}{Be}}}
\frac{F^{^{22}Na}}
     {F^{^{24}Na,^{7}Be}}\;\; 
\times \nonumber \\ 
\displaystyle
&\times& \; \displaystyle  
\frac{1+\displaystyle \left(\overline{\sigma}^{^{22}Na}_{n,x}\Phi_n /
\displaystyle \sigma^{^{22}Na}_{p,x}\Phi_p\right)}
{1+\left({\displaystyle \overline{\sigma}^{^{24}Na,^{7}{Be}}_{n,x}\Phi_n} /
{\displaystyle \sigma^{^{24}Na,^{7}{Be}}_{p,x}\Phi_p}\right)} \mbox{ .}
\end{eqnarray} 

Since the $\Phi_n/\Phi_p$ ratio does not exceed $\sim 2\%$ at any 
proton beam energy (see Fig.~\ref{fig3}), the following simplified 
formula can be used:

\begin{eqnarray}
\label{mon2}
\frac{\sigma^{^{24}{Na},^{7}{Be}}}
     {\sigma^{^{22}{Na}}}
&=&
\frac{A_0^{^{24}Na,^7Be}}{A_0^{^{22}Na}}
\frac{\left(\eta \varepsilon \right)^{^{22}Na}}
     {\left(
\eta \varepsilon \right)^{^{24}Na,^{7}{Be}}}
\frac{F^{^{22}Na}}
     {F^{^{24}Na,^{7}Be}} \mbox{ .}
\end{eqnarray} 	

\begin{figure}[t]
\vspace{1cm}
\centerline{\epsfxsize 8.0cm \epsffile{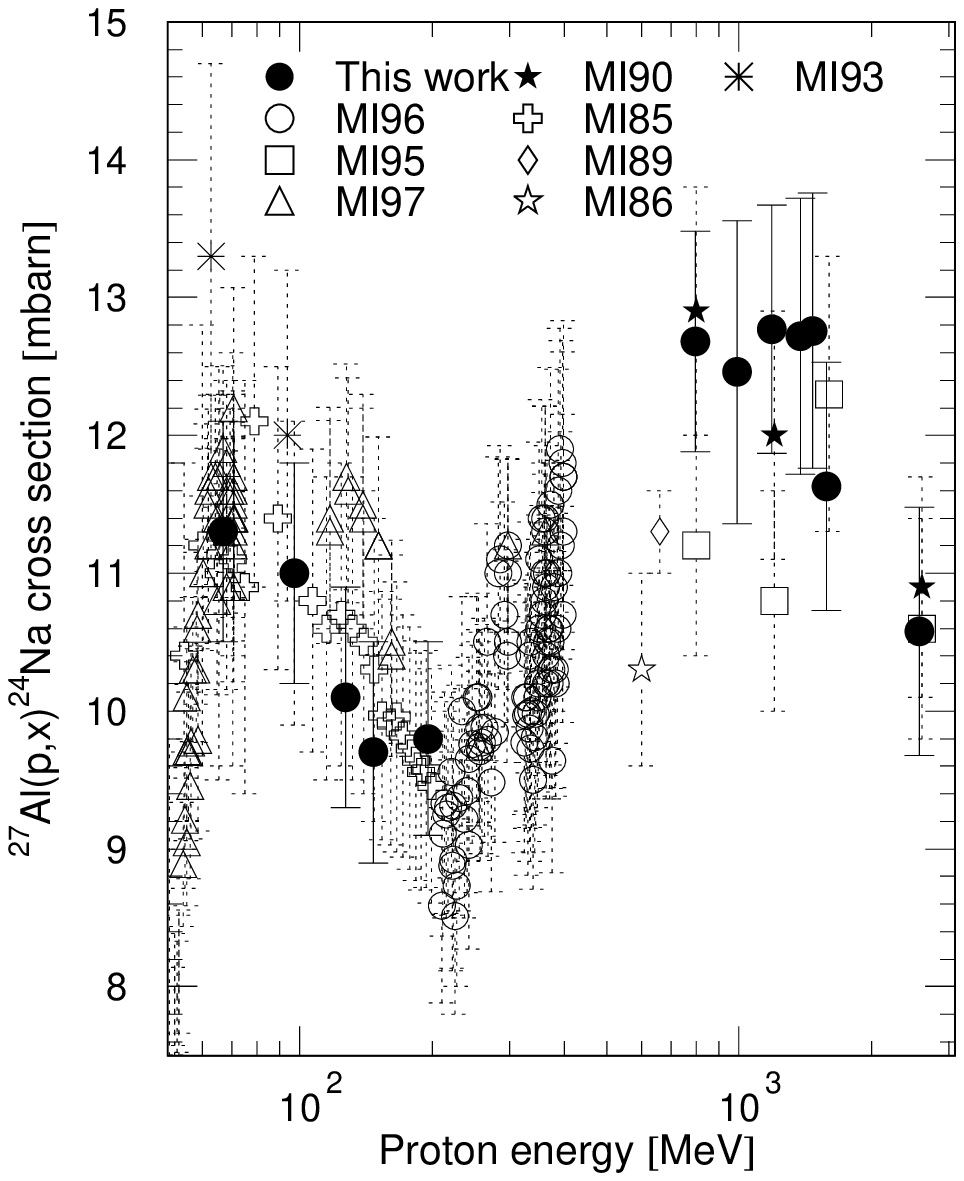}} 
\vspace*{2mm}
\caption{
The $^{27}$Al(p,x)$^{24}$Na monitor cross sections measured 
in this work and from previous works:
MI85 -- 
%\cite{mi85},
[18],
MI86 -- 
%\cite{mi86},
[19],
MI89 -- 
%\cite{mi89},
[20],
MI90 -- 
%\cite{mi90},
[21],
MI93 -- 
%\cite{mi93},
[22]
MI95 -- 
%\cite{7},
[7]
MI96 -- 
%\cite{mi96},
[23],
MI97 -- 
%\cite{mi97}.
[24].
}
\label{fig4}
\end{figure}

The urgent necessity for high-precision monitoring has made us to extend 
our measurements to the entire energy range used in our experiments, 
although a single energy (1 GeV) is treated in the present work.
Table \ref{crosssec} and Fig. \ref{fig4} show the measurement results.

\begin{table*}
\begin{center}
\caption{The monitor reaction cross sections [mb] averaged over the experiments}
\label{crosssec}
\vspace*{0.5cm}
\begin{tabular}{llll}
Proton energy, &The $^{27}${Al}(p,x)$^{22}${Na}&\multicolumn{2}{|c|}{The measured reaction cross sections.}\\
GeV            &monitor cross                  &\multicolumn{2}{|c|}{Shown in brackets are the errors disregarding/allowing}\\
                & sections used in &\multicolumn{2}{|c|}{for the $^{27}${Al}(p,x)$^{22}${Na} reaction cross section error}\\ \cline{3-4}
                & this work             & $^{27}${Al}(p,x)$^{24}${Na}& 		$^{27}${Al}(p,x)$^{7}${Be}\\ \hline
0.067	&	24.4 $\pm$ 1.4	&	11.3 $\pm$ ( 0.5 / 0.8 )	&	0.76 $\pm$ ( 0.20 / 0.21 ) \\ \hline
0.097	&	19.1 $\pm$ 1.3	&	11.0 $\pm$ ( 0.3 / 0.8 )	&	0.97 $\pm$ ( 0.07 / 0.10 )	\\ \hline
0.127	&	17.0 $\pm$ 1.3	&	10.1 $\pm$ ( 0.3 / 0.8 )	&	1.14 $\pm$ ( 0.06 / 0.11 )	\\ \hline
0.147	&	16.1 $\pm$ 1.2	&	 9.7 $\pm$ ( 0.4 / 0.8 )	&	1.44 $\pm$ ( 0.11 / 0.16 )	\\ \hline
0.197	&	15.1 $\pm$ 0.9	&	 9.8 $\pm$ ( 0.4 / 0.7 )	&	1.48 $\pm$ ( 0.04 / 0.10 )	\\ \hline
0.8	&	15.5 $\pm$ 0.9	&	12.7 $\pm$ ( 0.3 / 0.8 )	&	6.3 $\pm$ ( 0.3 / 0.4 )	\\ \hline
1.0	&	15.0 $\pm$ 0.9	&	12.5 $\pm$ ( 0.8 / 1.1 )	&	7.5 $\pm$ ( 0.3 / 0.5 )	\\ \hline
1.2	&	14.6 $\pm$ 1.0	&	12.8 $\pm$ ( 0.3 / 0.9 )	&	8.3 $\pm$ ( 0.2 / 0.6 )	\\ \hline
1.4	&	13.9 $\pm$ 1.0	&	12.7 $\pm$ ( 0.4 / 1.0 )	&	8.9 $\pm$ ( 0.3 / 0.7 )	\\ \hline
1.5	&	13.5 $\pm$ 1.0	&	12.8 $\pm$ ( 0.3 / 1.0 )	&	8.8 $\pm$ ( 0.3 / 0.7 )	\\ \hline
1.6	&	13.2 $\pm$ 1.0	&	11.6 $\pm$ ( 0.3 / 0.9 )	&	8.9 $\pm$ ( 0.2 / 0.7 )	\\ \hline
2.6	&	11.7 $\pm$ 0.9	&	10.6 $\pm$ ( 0.3 / 0.9 )	&	9.2 $\pm$ ( 0.2 / 0.7 )	\\ 
\end{tabular} 
\end{center}
\end{table*}

\subsection{
$\gamma$-spectrometer detection efficiency allowing for 
the height factors}

The $\gamma$ spectra have to be measured for a broad range of heights H 
(the source-detector distances) so that early counting may begin without 
overwhelmingly large counting rates, and later measurements do not suffer 
from too low rates.

Therefore, the $\gamma$ spectrometer was efficiency calibrated in a 
60 $\!<$ E$_{\gamma}$ $\!\!\!<$ 2600 keV range for nine 
heights within a 40--550 mm interval using the standard set of 
$\gamma$-radiation sources ($^{54}${Mn}, $^{57}${C­}, 
$^{60}${Co}, $^{88}${Y}, $^{109}${Cd}, $^{113}${Sn}, 
$^{133}${Ba}, $^{137}${Cs}, $^{139}${Ce}, 
$^{152}${Eu}, $^{228}${Th}, $^{241}${Am}, and $^{22}${Na}). The energy 
dependence of the detection efficiency obtained for 
a base height
was approximated by the cubic spline
\begin{equation}
\label{eff}
\varepsilon_{base}(E) = exp \left[ \sum _{i=k}^{k+3}P_i\cdot \left(\ln E 
\right)^{i-k} \right] , 
\end{equation}
\begin{eqnarray}
\label{deleff} 
&\displaystyle \Delta \varepsilon_{base} &= \varepsilon_{base}(E) 
\times \nonumber \\ 
&\times& \!\!\!\!\!\!\!\!\sqrt{ \sum^{k+3}_{i=k} 
\sum^{k+3}_{j=k} M^{-1}_{i j} (\ln{E})^{\;i+j-2k}} \cdot 
\sqrt{ \frac{\chi^2_{\varepsilon}}{L}} ,
\end{eqnarray}
where 
$$
k = 
\cases{1, & for $\ln E < \ln E_0$, \hspace*{4mm} $E_0 = 300$ keV,\cr
 5,  & for $\ln E \ge \ln E_0$, \cr} 
$$
\begin{equation}
\label{chisq}
\chi ^2_{\varepsilon} = \sum ^{J_1+J_2}_{i=1}\left[\frac{\varepsilon 
^{exp}_{base}(E_i) - \varepsilon_{base} \left(E_i \right)}{\Delta 
\varepsilon ^{exp}_{base}(E_i)} \right]^2,
\end{equation}
where
$\varepsilon_{base}\left(E_i \right)$ and
$\Delta \varepsilon ^{exp}_{base}(E_i)$ are the value and its absolute error 
for the calculated efficiency of the spectrometer at the i-th energy $E_i$; 
$P_i$ are polynomial coefficients;
$M^{-1}_{i j}$  is the covariance matrix of the polynomial coefficients;
$L$ is the number of degrees of freedom when determining  $\chi^2$; 
$J_1$ + $J_2$ is the total number of experimental points; 
$\varepsilon ^{exp}_{base}(E_i)$ and
$\Delta \varepsilon ^{exp}_{base}(E_i)$ are the experimental 
values and their absolute errors for the experimental efficiency 
of the spectrometer at the i-th energy $E_i$.

The radioactive reaction products of as long a half-life as possible were 
sought for and identified for each of the irradiated samples. Therefore, 
starting from a certain moment, the measurements were made at the lowest 
height (H=40 mm), where the effect of the cascade summation of 
$\gamma$ quanta is negligible. The magnitude of this effect was determined 
experimentally.

An optimal schedule for changing the heights to preserve an approximate 
constancy of the maximum possible spectrometer load was very difficult to 
select beginning from the very start moment of measuring a sample irradiated.
Therefore, the height dependence of the detection efficiency was analyzed for
each of the $\gamma$ energies of the calibration sources.
The analysis has given the dependence expressed as
\begin{equation}
\label{other}
\varepsilon \left( E, H \right) = \varepsilon_{base} 
\left( E \right) \cdot \!\!\left[\frac{\left(q_1+ q_2\cdot 
\ln E + H_{base} \right)}
{\left( q_1 + q_2\cdot \ln E + H \right)} \right]^2 ,
\end{equation}
\begin{eqnarray}
\label{de}
\Delta {\displaystyle\varepsilon}(E,H) = \varepsilon \left( E, H \right)  
\sqrt{\left[\frac{\Delta {{\displaystyle\varepsilon}_{base}}}
{\varepsilon_{base}}\right] ^2\!\!\!+\!4(H\!-\!H_{base})^2 \times } 
\nonumber
\\ 
\nonumber
\\
\overline{\times
\frac{(V^{-1}_{11} + 2 V^{-1}_{12}\cdot \ln E + V^{-1}_{22}\cdot (\ln E)^2) 
\cdot \chi _q^2} {\left[q_1 + 
q_2\cdot \ln E + H_{base} \right]^2 \cdot \left[q_1 + 
q_2\cdot \ln E + H \right]^2}} \mbox{ ,}
\end{eqnarray}
where $\varepsilon \left( E, H \right)$ and $\Delta 
{\displaystyle\varepsilon}(E,H)$ 
are, respectively, the value and the absolute error for the 
dependence of the spectrometer detection efficiency on energy 
and height; $q_1$ and $q_2$ are the fitting parameters; $V^{-1}_{ij}$ 
are the covariance matrix elements of the parameters $q_1$ and $q_2$; 
$\chi^2_q$ is given by Eq.~(\ref{chisq}). 

\begin{figure}[t] 
\centerline{\epsfxsize 8.0cm \epsffile{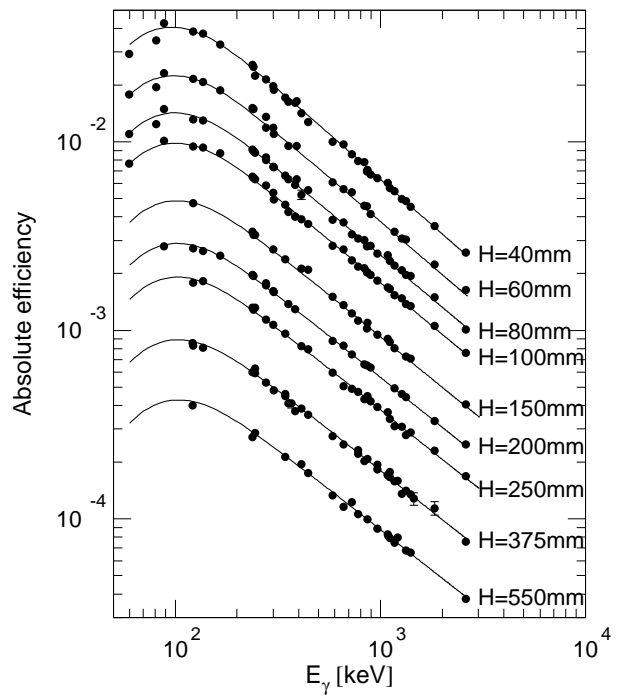}} 
\vspace*{2mm}
\caption{The measured and simulated 
height-energy detection 
efficiency of the CANBERRA GC2518 Ge-detector-based 
$\gamma$ spectrometer for different heights of the $\gamma$ source.}
\label{fig5}
\end{figure}

From formula (\ref{other}) it is readily seen that any of the above 
mentioned heights can be selected, but the extreme heights are not 
desirable, for they involve a rise of the systematic 
error when renormalizing the rest of the heights to the basic height.
If, however, the height of the prevailing measurements is selected, 
the systematic error decreases during the renormalization.

Fig.~\ref{fig5} shows the detection efficiencies measured at different 
heights (the dots). The curves in the figure present the dependence of 
the efficiencies on height and energy as inferred from formula (\ref{other}).

The analysis based on formulas (\ref{deleff}) and (\ref{de}) has shown that 
the systematic errors due to uncertainties in the detection efficiency do 
not exceed 5\%.

\subsection{Processing of $\gamma$ spectra and cross section determination}

The measured $\gamma$ spectra of irradiated $^{208}${Pb} are extremely 
complicated due to the numerous $\gamma$ lines. Despite using the Ge 
detector with its ultimate attained resolution, the spectra still include 
many multiplets, so the GENIE-2000 code \cite{genie} which is designed to
process 
complicated $\gamma$ spectra has been used. 
The code is notable for permitting 
the results of tentative processing to be analyzed again after the 
computer-aided consolidated processing of the $\gamma$ spectra by 
interactive fitting of the $\gamma$ peaks. Namely, we find out 
whether the peak regions are either multiplets, real peaks that fail to 
satisfy the search criteria, spurious peaks, etc.

The interactive processing 
has made it possible to improve the 
accuracy and reliability of analysis of the measured $\gamma$ spectra, 
especially in the case of poorly resolved spectra with poor statistics.

The processed $\gamma$ spectra are then united to form the input file 
for the ITEP-developed SIGMA code, which plots the time variations of a 
selected $\gamma$-line intensity and makes use of the energy and the 
half-life of that $\gamma$ line to identify the product nuclide using
the PCNUDAT database \cite{pcnud} and the Firestone's Table of Isotopes
\cite{fire} and to determine the product 
nuclide cross sections using formulas (\ref{1})--(\ref{3}).

\section{Measured product yields from the 
$^{208}${Pb}(p,x) reaction}

Table II
presents the results of determining the reaction
product yields in
1 GeV proton-irradiated $^{208}${Pb}. Out of 114 yields presented, 
8 are independent yields of ground states(i), 
15 are independent yields of metastable states (i($\Sigma m_j$)), 
15 are independent yields of metastable and ground states (i($\Sigma m_j$+g)), 
65 are cumulative yields (c), and 11 are supra cumulative 
yields, when the correction may exceed the determination error (c$^*$).
It should be noted that  
in case the (c) values are to be reconstructed from (c$^*$) using of 
formula (15), we must have the $\sigma_1^{cum}$ values that have 
yet to be determined, but are expected to 
be measured in the future.
As mentioned above, however, the simulation codes can be tested using the 
(c$^*$) values.

\begin{table*}  
\label{results}

\caption{Experimental product nuclide yields [mb] in 1 GeV 
proton-irradiated $^{208}${Pb} compared with both recent
GSI data measured in inverse kinematics [8] and the ZSR measurements 
on $^{nat}$Pb [29]}

\vspace*{0.5cm}

\begin{tabular}{|c|c|c|c|c|c|}      
{Product} & T$_{1/2} \cite{pcnud}, \cite{fire}$ & Type & Yield (this work) & 
Work \cite{8} & Work \cite{Michel}  \\ \hline
 \hline
\nuc{206}{Bi}&  6.243d  &      i     &      4.60 $\pm$     0.29 & -- & 5.36 $\pm$ 0.67 \\  
\nuc{205}{Bi}& 15.31d  &      i     &      6.20 $\pm$     0.40 & -- & 7.09 $\pm$ 0.90 \\ 
\nuc{204}{Bi}& 11.22h  &      i(m1+m2+g)     &      5.29 $\pm$     0.80 & -- & 6.03  $\pm$ 0.95 \\ 
\nuc{203}{Bi}& 11.76h  &      i(m+g)     &      4.84 $\pm$     0.59 & -- &  -- \\ 
\nuc{204m}{Pb}& 67.2m  &      i(m)   &      11.0 $\pm$      1.0 & -- &  -- \\ 
\nuc{203}{Pb}& 51.873h  &      c     &      31.5 $\pm$      2.1 & 28.7  $\pm$ 3.1 &  -- \\ 
\nuc{201}{Pb}&  9.33h  &  c$^{*}$         &      26.9 $\pm$      2.4 & 20.4  $\pm$ 1.9  &  -- \\ 
\nuc{200}{Pb}& 21.5h  &  c   &      18.2 $\pm$      1.2 & 18.2  $\pm$ 2.0 & 27.8  $\pm$ 3.5 \\ 
\nuc{198}{Pb}&  2.4h  &  c       &       8.9 $\pm$      2.1 & 14.0  $\pm$ 1.3 &  -- \\ 
\nuc{197m}{Pb}& 43m  &  c$^{*}$            &      17.9 $\pm$      4.0 & -- &  -- \\ 
\nuc{202}{Tl}& 12.23d  &  c       &      18.8 $\pm$      1.2 & 40.0  $\pm$ 4.0 & 22.0  $\pm$ 2.7 \\ 
\nuc{201}{Tl}& 72.912h  &  c           &      43.7 $\pm$      2.9 & 37.3  $\pm$ 3.7 & 53.5  $\pm$ 6.6\\ 
\nuc{200}{Tl}& 26.1h  &  c         &      40.6 $\pm$      2.6 & 35.2  $\pm$ 3.7 &  -- \\ 
\nuc{200}{Tl}& 26.1h  &  i(m+g)         &      22.7 $\pm$      1.5 &17.0  $\pm$ 1.7  & 22.3  $\pm$ 6.1 \\ 
\nuc{199}{Tl}&  7.42h  &  c         &      38.5 $\pm$      5.2 & 34.3  $\pm$ 3.4 &  -- \\ 
\nuc{198m1}{Tl}&  1.87h  &  i(m1+m2)     &      17.6 $\pm$      3.6 & -- &  -- \\ 
\nuc{198}{Tl}&  5.3h  &  c         &      35.9 $\pm$      5.0 & --  &  -- \\ 
\nuc{196m}{Tl}& 1.41h & i(m)         &      34.8 $\pm$      4.4 & -- &  -- \\ 
\nuc{203}{Hg}& 46.612d  &  c         &      4.03 $\pm$     0.27 & -- &3.66   $\pm$ 0.45 \\ 
\nuc{197m}{Hg}& 23.8h  &  i(m)         &      10.7 $\pm$      0.7 & -- &  -- \\ 
\nuc{195m}{Hg}& 41.6h  &  i(m)         &      13.6 $\pm$      2.0 & -- & 13.3  $\pm$ 1.8 \\ 
\nuc{193m}{Hg}& 11.8h  &  i(m)         &      18.9 $\pm$      2.5 & -- & 10.8  $\pm$ 2.3 \\ 
\nuc{192}{Hg}&  4.85h  &  c         &      35.2 $\pm$      2.8 &31.3  $\pm$ 3.4  &  -- \\ 
\nuc{198m}{Au}& 2.27d  &  i(m)           &      1.01 $\pm$     0.14 & -- & 1.25  $\pm$ 1.11 \\ 
\nuc{198}{Au}& 2.69517d  &  i(m+g)         &      2.11 $\pm$     0.22 & 1.96  $\pm$ 0.23 &  -- \\ 
\nuc{198}{Au}& 2.69517d  &  i           &      1.09 $\pm$     0.30 & -- &  -- \\  
\nuc{196}{Au}&  6.183d  &  i(m1+m2+g)  &      4.13 $\pm$     0.35 & 4.02  $\pm$ 0.47 & 3.88  $\pm$ 0.47 \\ %(G) \\ 
\nuc{195}{Au}& 186.098d &  c           &      48.7 $\pm$      5.5 & 28.4  $\pm$ 3.3  & 51.1  $\pm$ 6.6 \\ %(G) \\ 
\nuc{194}{Au}& 38.02h  &  i(m1+m2+g) &      7.06 $\pm$     0.75 & 6.33  $\pm$ 0.75 & 6.85  $\pm$ 0.92 \\ 
\nuc{192}{Au}&  4.94h  &  c           &      46.9 $\pm$      6.6 & 39.9  $\pm$ 4.6 &  -- \\ 
\nuc{192}{Au}&  4.94h  &  i(m1+m2+g) &      11.6 $\pm$      1.7 & 9.2   $\pm$ 1.1 &  -- \\ 
\nuc{191}{Pt}& 2.802d  &  c     &      41.8 $\pm$      4.2 & 44.4  $\pm$ 5.5 & 37.9  $\pm$ 4.8 \\ 
\nuc{189}{Pt}& 10.87h  &  c         &      46.8 $\pm$      4.8 &40.4  $\pm$ 5.0  &  -- \\ 
\nuc{188}{Pt}& 10.2d  &  c     &      40.5 $\pm$      2.9 & 38.4  $\pm$ 4.7 & 42.8  $\pm$ 5.4 \\ 
\nuc{186}{Pt}& 2.08h &  c$^{*}$ & 33.5  $\pm$  2.3 & 32.9 $\pm$   4.1 & -- \\ 
\nuc{190}{Ir}& 11.78d  &  i(m1+g)         &      0.69 $\pm$     0.06 & -- &  -- \\ 
\nuc{188}{Ir}& 41.5h  &  c     &      43.2 $\pm$      3.2 & 40.9  $\pm$ 5.4 &  -- \\ 
\nuc{188}{Ir}& 41.5h  &   i  &      2.93 $\pm$     0.69 & 2.48  $\pm$ 0.33 &  -- \\ 
\nuc{186}{Ir }& 16.64h  &  i         &      20.8 $\pm$      1.9 & -- & 22.5  $\pm$ 3.1 \\ %(G) \\ 
\nuc{185}{Ir}& 14.4h  &  c$^{*}$         &      34.8 $\pm$      2.3 & 39.4  $\pm$ 5.2  & 39.4  $\pm$ 7.9 \\ 
\nuc{184}{Ir}&  3.09h  &  c$^{*}$         &      39.5 $\pm$      3.0 & 36.9  $\pm$ 4.8 &  -- \\ 
\nuc{185}{Os}& 93.6d  &  c         &      41.8 $\pm$      2.8 & 38.1  $\pm$ 5.3  & 43.0  $\pm$ 5.3 \\ 
\nuc{183m}{Os}&  9.9h  &  c         &      23.2 $\pm$      1.5 & -- &  -- \\ 
\nuc{182}{Os}& 22.10h  &  c         &      42.0 $\pm$      2.8 & 34.2  $\pm$ 4.8 &  -- \\ 
\nuc{183}{Re}& 70.0d  &  c         &      41.7 $\pm$      2.9 & 36.3  $\pm$ 5.3 & 38.2  $\pm$ 4.8 \\ 
\nuc{182m}{Re}& 12.7h  &  c         &      45.2 $\pm$      3.7 & -- &  -- \\ 
\nuc{181}{Re}& 19.9h  &  c         &      43.1 $\pm$      5.9 & 37.0  $\pm$ 5.4 & 45.9  $\pm$ 5.9 \\ 
\nuc{179}{Re}& 19.5m  &  c$^{*}$&      48.2 $\pm$      4.2 & 44.7  $\pm$ 6.6 &  -- \\ 
\nuc{177}{W}&  135m  &  c         &      30.1 $\pm$      3.5 &  23.4  $\pm$ 3.6&  -- \\ 
\nuc{176}{W}&  2.5h  &  c         &      28.0 $\pm$      3.9 & 29.0  $\pm$ 4.5 &  -- \\ 
\nuc{176}{Ta}&  8.09h  &  c           &      35.0 $\pm$      3.6 & 28.8  $\pm$ 4.7 &  -- \\  
\nuc{173}{Ta}&  3.14h  &  c         &      30.9 $\pm$      3.9 & 26.3  $\pm$ 4.3 &  -- \\ 
\nuc{172}{Ta}& 36.8m  &  c$^{*}$&      17.3 $\pm$      2.3 & 27.4  $\pm$ 4.5  &  -- \\ 
\nuc{175}{Hf}& 70d  &  c         &      31.3 $\pm$      2.3 & 28.3  $\pm$ 4.8 & 34.1  $\pm$ 4.1 \\ 
\nuc{173}{Hf}& 23.6h  &  c         &      28.4 $\pm$      2.6 & 25.2  $\pm$ 4.3 & 39.0  $\pm$ 4.9 \\ 
\nuc{172}{Hf}& 1.87y &  c         &      24.1 $\pm$      1.6 & 24.6  $\pm$ 4.2 & 24.4  $\pm$ 3.1 \\        
   \end{tabular}

\end{table*}

\begin{table*} 

  \begin{tabular}{|c|c|c|c|c|c|}      
{Product} & T$_{1/2}$ & Type & Yield (this work) & 
Work \cite{8} & Work \cite{Michel}  \\ \hline
\hline
\nuc{171}{Hf}& 12.1h  &  c         &      18.2 $\pm$      2.8 & 22.9  $\pm$ 3.9 &  -- \\ 
\nuc{170}{Hf}& 16.01h  &  c         &      22.1 $\pm$      6.8 & 20.3 $\pm$ 3.5 & 21.2  $\pm$ 3.0 \\ 
\nuc{172}{Lu}&  6.70d  &  c         &      23.9 $\pm$      1.7 & 24.7  $\pm$ 4.4 &  -- \\ 
\nuc{172}{Lu}&  6.70d  &  i(m1+m2+g)  &      0.19 $\pm$     0.05 & 0.183  $\pm$ 0.037 &  -- \\ 
\nuc{171}{Lu}&  8.24d  &  c         &      26.1 $\pm$      1.8 & 16.6  $\pm$ 3.0 & 31.3  $\pm$ 3.9 \\ 
\nuc{170}{Lu}& 2.012d  &  c         &      21.7 $\pm$      2.9 & 20.9  $\pm$ 3.7 &  -- \\ 
\nuc{169}{Lu}& 34.06h  &  c         &      18.6 $\pm$      1.2 & 12.1  $\pm$ 2.2 & 26.4  $\pm$ 3.7 \\ %(G) \\ 
\nuc{169}{Yb}& 32.026d  &  c         &      20.9 $\pm$      1.5 & 18.1 $\pm$ 3.4 & 24.3  $\pm$ 3.0 \\ %(G) \\ 
\nuc{166}{Yb}& 56.7h  &  c         &      16.1 $\pm$      1.1 & 13.7  $\pm$ 2.6 & 16.4  $\pm$ 2.3 \\ 
\nuc{167}{Tm}&  9.25d  &  c         &      19.4 $\pm$      4.0 &14.0  $\pm$ 2.7  & 21.2  $\pm$ 2.6 \\ 
\nuc{165}{Tm}& 30.06h  &  c         &      14.4 $\pm$      1.4 & 13.3  $\pm$ 2.6 &  -- \\ 
\nuc{160}{Er}& 28.58h  &  c         &      8.8 $\pm$      0.6 & 7.2   $\pm$ 1.5  &  -- \\ 
\nuc{157}{Dy}&  8.14h  &  c         &      5.73 $\pm$     0.45 & 5.0   $\pm$ 1.1  &  -- \\ 
\nuc{155}{Dy}&  9.9h  &  c$^{*}$ &      3.66 $\pm$     0.27 & 2.86  $\pm$ 0.63 &  -- \\ 
\nuc{155}{Tb}&  5.32d  &  c         &      4.16 $\pm$     0.39 & 2.72  $\pm$ 0.62 & 5.52  $\pm$ 0.70 \\ 
\nuc{153}{Tb}& 2.34d  &  c$^{*}$ &      2.52 $\pm$     0.25 &2.40 $\pm$ 0.54  & 2.51  $\pm$ 0.40 \\ 
\nuc{152}{Tb}& 17.5h  &  c$^{*}$&      2.10 $\pm$     0.17 & -- &  -- \\ 
\nuc{153}{Gd}& 240.4d &  c         &      2.65 $\pm$     0.24 & 2.18 $\pm$ 0.51 &  3.10  $\pm$ 0.38 \\ 
\nuc{149}{Gd}&  9.28d  &  c         &      2.24 $\pm$     0.18 & -- & 3.06  $\pm$ 0.38    \\ 
\nuc{146}{Gd}& 48.27d  &  c         &      1.26 $\pm$     0.09 & 1.23  $\pm$ 0.29 & 1.68  $\pm$ 0.21 \\  
\nuc{147}{Eu}& 24.1d  &  c         &      0.98 $\pm$     0.30 & 1.18  $\pm$ 0.29 & 1.97  $\pm$ 0.29 \\ 
\nuc{146}{Eu}&  4.61d  &  c         &      1.62 $\pm$     0.12 & 1.17  $\pm$ 0.28 &  -- \\ 
\nuc{146}{Eu}&  4.61d  &  i         &      0.37 $\pm$     0.05 & 0.181  $\pm$ 0.047 &  -- \\ 
\nuc{143}{Pm}& 265d &  c         &      1.02 $\pm$     0.13 & 0.85  $\pm$ 0.22 & 1.00  $\pm$ 0.13 \\ 
\nuc{139}{Ce}& 137.640d &  c         &      0.83 $\pm$     0.06 & 0.45  $\pm$ 0.13 &  0.82  $\pm$ 0.10  \\ 
\nuc{121m}{Te}& 154d &  i(m)    &      0.44 $\pm$     0.04 & -- &  0.53  $\pm$ 0.07  \\ 
\nuc{121}{Te}& 19.16d  &  c         &      1.07 $\pm$     0.11 &  -- & 0.79  $\pm$ 0.10 \\ 
\nuc{119m}{Te}&  4.70d  &  i(m)        &      0.40 $\pm$     0.04 & -- &  -- \\ 
\nuc{120m}{Sb}&  5.76d  &  i(m)       &      0.54 $\pm$     0.05 & -- & 0.53   $\pm$  0.07 \\ 
\nuc{114m}{In}& 49.51d  &  i(m1+m2) &      0.95 $\pm$     0.19 & -- & 1.07  $\pm$ 0.16 \\% (M1) \\ 
\nuc{110m}{Ag}& 249.76d &  i(m)     &      1.11 $\pm$     0.09 & -- & 1.32  $\pm$ 0.17 \\ 
\nuc{106m}{Ag}&  8.28d  &  i(m)           &      0.89 $\pm$     0.08 & -- & 0.92  $\pm$ 0.14 \\ 
\nuc{105}{Ag}& 41.29d  &  c         &      0.65 $\pm$     0.12 & 0.74  $\pm$ 0.17 & 1.04  $\pm$ 0.14 \\ %(G) \\ 
\nuc{105}{Rh}& 35.36h  &  c           &      4.63 $\pm$     0.54 & 3.13  $\pm$ 0.51 &  -- \\ 
\nuc{101m}{Rh}&  4.34d  &  c           &      1.29 $\pm$     0.16 & -- &  -- \\ 
\nuc{103}{Ru}& 39.26d  &  c         &      3.84 $\pm$     0.26 & 3.03  $\pm$ 0.50 & 4.11  $\pm$ 0.53 \\ 
\nuc{96}{Tc}&  4.28d  &  i(m+g)   &      1.20 $\pm$     0.09 & -- & 1.49  $\pm$ 0.19 \\ 
\nuc{95}{Tc}& 20.0h  &  c         &      1.38 $\pm$     0.13 &  -- &  -- \\ 
\nuc{96}{Nb}& 23.35h  &  i         &      2.31 $\pm$     0.19 & 2.13  $\pm$ 0.34 &  -- \\ 
\nuc{95}{Nb}& 34.975d  &  c         &      5.41 $\pm$     0.34 & -- &  -- \\ 
\nuc{95}{Nb}& 34.975d  &  i(m+g)     &      3.03 $\pm$     0.20 & -- & 3.58   $\pm$ 0.56 \\ 
\nuc{95}{Zr}& 64.02d  &  c         &      2.34 $\pm$     0.15 & 1.58  $\pm$ 0.28 & 2.32  $\pm$ 0.29 \\ 
\nuc{89}{Zr}& 78.41h  &  c         &      2.30 $\pm$     0.16 &  -- & 2.82  $\pm$ 0.35 \\ %(G) \\ 
\nuc{88}{Zr}& 83.4d  &  c           &      0.76 $\pm$     0.08 & 0.97 $\pm$ 0.15 & 1.19  $\pm$ 0.15 \\ 
\nuc{90m}{Y}&  3.19h  &  i(m)      &      4.82 $\pm$     0.39 & -- &  -- \\ 
\nuc{88}{Y}& 106.65d &  c         &      4.03 $\pm$     0.27 & 3.72  $\pm$ 0.58 &  -- \\ 
\nuc{88}{Y}& 106.65d &  i(m+g)  &      3.41 $\pm$     0.25 &  2.76 $\pm$ 0.44 & 3.74  $\pm$ 0.46 \\ 
\nuc{87}{Y}& 79.8h  &  c$^{*}$&      2.94 $\pm$     0.23 & -- & 3.36  $\pm$ 0.42 \\ %(G) \\  
\nuc{85}{Sr}& 64.84d  &  c         &      2.76  $\pm$     0.22 & -- & 3.42  $\pm$ 0.41 \\ %(G) \\ 
\nuc{86}{Rb}& 18.631d  &  i(m+g)  &      5.48  $\pm$     0.66 & 2.43  $\pm$ 0.38 & 4.39  $\pm$ 0.61 \\ %(G) \\ 
\nuc{83}{Rb}& 86.2d  &  c         &      3.46  $\pm$     0.28 & 2.82  $\pm$ 0.45 & 3.96  $\pm$ 0.49 \\ 
\nuc{82m}{Rb}&  6.472h  &  i(m)       &      2.73  $\pm$     0.30 & -- &  -- \\ 
\nuc{82}{Br}& 35.30h  &  i(m+g)    &      2.17  $\pm$     0.14 & 1.55  $\pm$ 0.24 & 2.62  $\pm$ 0.50 \\ 
\nuc{75}{Se}& 119.779d &  c         &      1.33  $\pm$     0.09 & 1.18  $\pm$ 0.19 & 1.61  $\pm$ 0.20 \\
\nuc{74}{As}& 17.77d  &  i         &      1.86  $\pm$     0.18 & 1.66  $\pm$ 0.27 & 2.24  $\pm$ 0.28 \\ 
\nuc{59}{Fe}& 44.472d  &  c         &      0.91  $\pm$     0.08 & 0.69  $\pm$ 0.11 & 1.05  $\pm$ 0.14 \\ 
\nuc{65}{Zn}& 244.26d  &  c         &      0.79  $\pm$     0.19 & 0.42  $\pm$ 0.07 & 0.66  $\pm$ 0.17 \\ 
\nuc{46}{Sc}& 83.79d  &  i(m+g)  &      0.35  $\pm$     0.06 & -- &  0.37  $\pm$ 0.05\\        
   \end{tabular} 

\end{table*}

>From Table II
we see that the experimental errors range 
from 7\% to 30\%.
The tabulated values are calculated as follows. Since most of the results 
are obtained by averaging over a set of ($\sigma_i \pm \Delta\sigma_i$) 
values calculated from different $\gamma$ lines, the mean and the 
experimental errors were calculated as
\begin{equation}
\label{mean}
\overline{\sigma}=\frac{\sum_i \sigma_iW_i}{\sum_iW_i}                  
, \qquad \mbox{where} \qquad W_i=1/\Delta \sigma_i^2 \mbox{ .}
\end{equation}
The $\Delta \sigma_i$  were determined using the error propagation 
formulas \cite{huds}
\begin{equation}
\label{er1}
\Delta \overline{\sigma}^{\;\;'}=
\sqrt{\frac{\sum_iW_i\left(\overline{\sigma}-\sigma_i\right)^2}
{\left(n-1\right)\sum_iW_i}} \mbox{ ,}                                
\end{equation}
\begin{equation}
\label{er2}
\Delta \overline{\sigma}^{\;\;''}=
\sqrt{\frac{1}{\sum_i W_i}} \mbox{ .}                                
\end{equation}

In the calculations, the largest of $\Delta \overline{\sigma}^{\;\;'}$ 
and $\Delta \overline{\sigma}^{\;\;''}$ was taken to be the experimental 
error $\Delta \overline{\sigma}$. Allowing for the monitor error, the total 
error in the measured cross sections was calculated as
\begin{equation}
\label{total_er}
\frac{\Delta \overline{\sigma}}{\overline{\sigma}}=\sqrt{\left(
\frac{\Delta \overline{\sigma}}{\overline{\sigma}}\right)^2+
\left(\frac{\Delta {\sigma_{st}}}{{\sigma_{st}}}\right)^2 } \mbox{ .}              
\end{equation}
Our analysis has shown that the main contribution to the total error 
is from uncertainties in the nuclear data (the absolute gamma yields 
and cross sections of the monitor reactions).

\section{Comparison with experimental data obtained elsewhere}

In addition to the nuclide production cross sections measured 
in this work, 
Table II
also presents data from other works, namely, the 
results obtained by the ``inverse kinematics" technique 
(\nuc{208}{Pb} bombarding \nuc{1}{H} at GSI \cite{8})
and the results of measuring the nuclide production 
in 1 GeV proton-irradiated \nuc{nat}{Pb} obtained by a technique that is 
similar to ours 
by the group of R. Michel at 
Hannover University, Zentrum f\"{u}r Strahlenschutz und Radio\"{o}kologie
(ZSR)~\cite{Michel}.

It should be noted that the GSI data are measured
only as independent yields, so 
they are summed up for their isobaric chains to be compared with the 
respective cumulative yields determined at ITEP and ZSR.
We estimated errors for the GSI cumulative yields as square roots of
sums of squared errors of all independent yields contributing to the
corresponding cumulative cross sections and also took into account
their systematic errors which vary from 9 to 25\% \cite{8}.
The GSI technique does not distinguish if a product nuclide is in 
a ground or metastable state, hence, we do not compare
our data with the GSI set in the case we measured 
only
metastable or ground states
(\nuc{204m}{Pb}, \nuc{197m}{Pb}, \nuc{198m1}{Tl}, \nuc{196m}{Tl}, \nuc{197m}{Hg}, 
\nuc{195m}{Hg}, \nuc{193m}{Hg}, \nuc{198m}{Au}, \nuc{198g}{Au}, \nuc{186g}{Ir}, 
\nuc{183m}{Os}, \nuc{182m}{Re}, \nuc{121m}{Te}, \nuc{119m}{Te}, \nuc{120m}{Sb}, 
\nuc{114m}{In}, \nuc{110m}{Ag}, \nuc{106m}{Ag}, \nuc{101m}{Rh}, \nuc{90m}{Y}, 
and \nuc{82m}{Rb})
as well as when  
there is a transition of a metastable state to a product out of 
the given decay chain
(\nuc{198}{Tl}, \nuc{190}{Ir}, \nuc{152}{Tb}, 
\nuc{149}{Gd},
\nuc{121}{Te}, \nuc{96}{Tc}, \nuc{95}{Tc}, \nuc{95}{Nb}, \nuc{95}{Nb}, 
\nuc{89}{Zr}, \nuc{87}{Y}, and \nuc{85}{Sr}).

Let us mention here
that the ZSR work  \cite{Michel} was carried out using 
natural lead, rather than high \nuc{208}{Pb}-enriched lead used at 
ITEP and GSI. We think, it is this fact that can be primarily responsible 
for some divergences between the ZSR data and the GSI and ITEP ones. The 
importance of this circumstance is indirectly confirmed by a 1.5 times 
higher mean-squared deviation of the ZSR data from the GSI data compared 
with the mean-squared deviation of the ITEP data from the same GSI data.
The mean-squared deviations 
between two data sets (1 and 2)
calculated as 
 $<F_{1-2}> = 100\% \times (10^{\sqrt{D}}-1)$ 
(where $D=<log(\sigma_{1i}/\sigma_{2i})^2>$ ) proved to be 
48\% and 32\% for ZSR--GSI and ITEP--GSI comparisons, respectively.

\section{Simulation of experimental data}

The present work is aimed at determining the nuclear cross sections to be 
used in designing ADS facilities. An attempt to obtain the necessary nuclear 
cross sections only from  experiments would involve impractical levels of 
expense 
and effort. Therefore, simulation techniques must be used for that purpose.
The simulation approach has the advantage that it also can be used for many 
other purposes. On the other hand, the current accuracy and reliability 
of simulated data are inferior to experiment. Also existing simulation 
codes have different predictive abilities when used to study the reactions 
that are of practical importance.

Considering this, the present work is aimed also at verifying the 
simulation codes used most extensively for this purpose in order to 
estimate their predictive abilities and to stimulate efforts 
to improve them.

The following seven simulation codes were examined to meet these 
requirements:
\begin{itemize}
\item  
the CEM95 cascade-exciton model code \cite{15},
\item  
the latest version of the improved cascade-exciton model \cite{cem97}
code, CEM2k, \cite{cem2k}, 
\item  
the CASCADE cascade-evaporation-fission transport code \cite{16},
\item  
the INUCL cascade-preequilibrium-evaporation-fission code \cite{17},
\item  
the LAHET (both ISABEL and Bertini options)
cascade-preequilibrium-evaporation-fission transport
code \cite{19},
\item  
the YIELDX semi-phenomenological code \cite{20},
\item  
the CASCADE/INPE cascade-evaporation-preequilibrium-fission-transport 
code \cite{21}.
\end{itemize}

All the codes, except for CASCADE/INPE and CEM2k are described in some 
detail in \cite{6}.

The intranuclear part of the CASCADE/INPE code package \cite{21},
based on the Dubna cascade model \cite{22}, 
is used to simulate 
the characteristics of projectile interactions with target nuclei. Recently 
the code was upgraded at the Institute of Nuclear Power Engineering (Obninsk, 
Russia) \cite{21}. Principal modifications of the code include:
\begin{itemize}
\item 
a special routine for calculating the precompound spectra of 
$\alpha$ particles, which have been demonstrated to play an important role 
in the production of long-lived radioactivity in different heavy targets;
\item  
the code was upgraded to allow for the description of angular distributions 
of nucleons based on the Kalbach parametrization for projectile energies 
below 0.8 GeV;
\item  
a new realization of the Weisskopf evaporation approach used in
\cite{22} has been introduced.
This is based on inverse-reaction cross-section calculations 
using various optical potentials appropriate for various mass and energy 
regions; nuclear level densities are calculated taking into consideration 
nucleon pairing to evaluate the excitation energy of residual nuclei;
\item  
the latest version of the mass table of the nuclides \cite{23} has been used 
for binding-energy calculations.
\end{itemize}

The improved Cascade-Exciton Model (CEM) of nuclear reactions \cite{cem97}
was developed and incorporated in the code CEM97 
in the Theoretical Division of Los 
Alamos National Laboratory, as an improvement to the code CEM95 \cite{15}.
It is described in detail in \cite{cem97}, therefore we will not elaborate 
here. CEM2k is a next step in the improvement of the CEM;
it differs from CEM97 mainly in the details of the transitions from the
cascade stage of a reaction to the preequilibrium one, and from 
the latter to  equilibrium decay. This preliminary version of
CEM2k has less preequilibrium emission than the earlier versions. The changes
were motivated by discrepancies with the recent GSI data \cite{8} 
in the earlier versions of the model. It is 
briefly surveyed in \cite{cem2k}, it is
still under development and 
will be described in a future paper.

Contrary to the simulated data, the experimental results include not only 
independent, but also (and mainly) cumulative and supra cumulative, residual 
product nuclei. To get a correct comparison between the experimental and 
simulation data, the cumulative yields must be calculated on the basis of 
the simulated independent yields. If the production chain of $n$ radioactive 
nuclei is presented as
$$
\begin{array}{ccccccc} \label{ret}
\sigma_1 &  & \sigma_2 & & ... & & \sigma_n\\
\downarrow & & \downarrow &  & & & \downarrow  \\
1 & \stackrel{\displaystyle \nu_1}\longrightarrow & 2 & 
\stackrel{\displaystyle \nu_2}\longrightarrow & ... & 
\stackrel{\displaystyle \nu_{n-1}}\longrightarrow & n \\ 
\end{array} \eqno (33)
$$
\addtocounter{equation}{1}
(where $\nu_1 ,..., \nu_{n-1}$ are the branching ratios of the respective 
nuclides), the simulated cumulative and supra cumulative yields of the 
$n$-th nuclide can be calculated as 
\begin{equation}
\label{branch1}
\sigma^{cum}_n =  \sigma^{ind}_n + \sum^{n-1}_{i=1} 
\left( \sigma^{ind}_i \prod^{n-1}_{j=i} \nu_j \right) \mbox{,}
\end{equation}
\begin{equation}
\label{branch2} \begin{array}{l}
\displaystyle \sigma_n^{cum^{\displaystyle*}} = \sigma_n^{ind} + 
\frac{\lambda_{n-1}}{\lambda_{n-1} - \lambda_{n}} 
\nu_{n-1} \\  \displaystyle
\times \left[\sigma_{n-1}^{ind} + \sum^{n-2}_{i=1} 
\left( \sigma_i^{ind} \prod^{n-2}_{j=i} \nu_j \right)\right] \end{array} 
\mbox{ .}
\end{equation}

The branching ratios of the decay chains were taken from \cite{fire}, 
considering that the branched (due to isomeric transitions and 
$\alpha$ decay) isobaric chains can always be presented 
as a superposition of linear chains.

To get a correct comparison between results obtained by 
different codes, the calculations 
were renormalized to the same cross sections 
for proton-nucleus inelastic interactions. We calculated these cross 
sections by the semi-empirical formula \cite{25}
\begin{eqnarray}
\label{letaw}
\sigma_{inel} &=& 45 A^{0.7} f(A) g(E), (\mbox{ in mb}) \\			
f(A) &=& 1+0.016 \sin (5.3-2.63 \log A) , 
\nonumber
\\
g(E) &=& 1-0.62 \exp (-E/200) \sin (10.9E-0.28) ,
\nonumber
\end{eqnarray}
where $A$ is the mass number of the target
and $E$ is the energy  (in MeV) of the projectile proton.

The cross section for the $p +  ^{208}${Pb} inelastic interactions 
as calculated by formula (\ref{letaw}) is of 1857.8 mb at 995 MeV, 
the incident energy of our experiment.

If an experiment-simulation difference of no more than
30\% (0.77$<\sigma_{calc}/\sigma_{exp}<$1.3) 
is taken to be the coincidence criterion \cite{26}, 
the simulation accuracy can be defined as the ratio of the number 
of such coincidences to the number of the comparison events.
The 30\% level meets the accuracy requirements 
of cross sections for nuclide production to be used in designing 
ADS plants, according to \cite{26}. The mean 
simulated-to-experimental data ratio can be used as another coincidence 
criterion \cite{6}:
\begin{equation}
\label{coin}
<F> \; = 10^{\sqrt{\displaystyle <\log \left(\sigma_{cal,i}/
\sigma_{exp,i}\right)^2>}} ,
\end{equation}
with its standard deviation
\begin{equation}
\label{deviat}
S(<F>) = \; <\left(\log\left(\sigma_{cal, i}/\sigma_{exp, i}\right) - 
\log(<F>)\right)^2> ,
\end{equation}
where $<>$  designates averaging over all the experimental and 
simulated results used in the comparisons ($i = 1, ... , N_S$).

The mean ratio $<F>$ together with its standard deviation $S(<F>)$ defines 
the interval [$<F>/S(<F>)$ , $<F>\times S(<F>)$] that covers about 2/3 of 
the simulation-to-experiment ratios. A logarithmic scale is 
preferable when determining the factor $<F>$ rather than a linear scale, 
because the simulation-experiment differences may be as high as a few 
orders of magnitude.

We apply the above two criteria together with our
results shown in Figs.~\ref{fig6}--\ref{fig8},
to infer conclusions about the predictive power 
of a given code.

The default options were used in all of the simulation codes without 
modifying the codes to get optimal agreement  with the
data. All the calculations were made before any experimental results
were obtained, except the results from CEM2k. With such an approach, 
our comparisons demonstrate the real predictive power, rather than the
descriptive power of the codes.

\section{Comparison of experiment with simulations}

\begin{center}
\begin{figure*} 
\includegraphics[angle=-90,width=17.5cm]{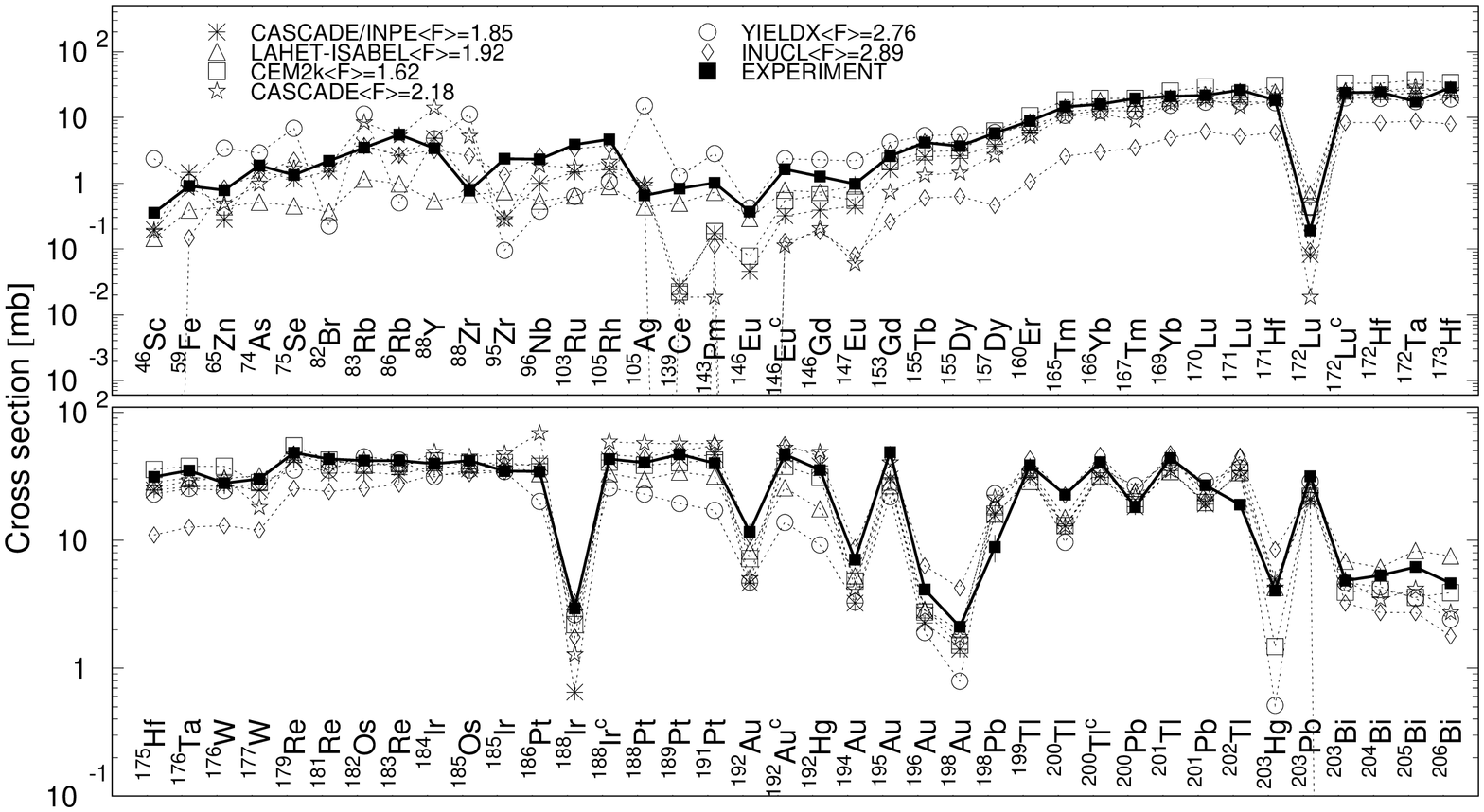}
\vspace*{2mm}
\caption{Detailed comparison between experimental and simulated yields
of radioactive reaction products. The cumulative yields are labeled 
with a ``c"
when the respective independent yields are also shown.
}
\label{fig6}
\vspace*{3mm}
\includegraphics[angle=-90,width=17.7cm]{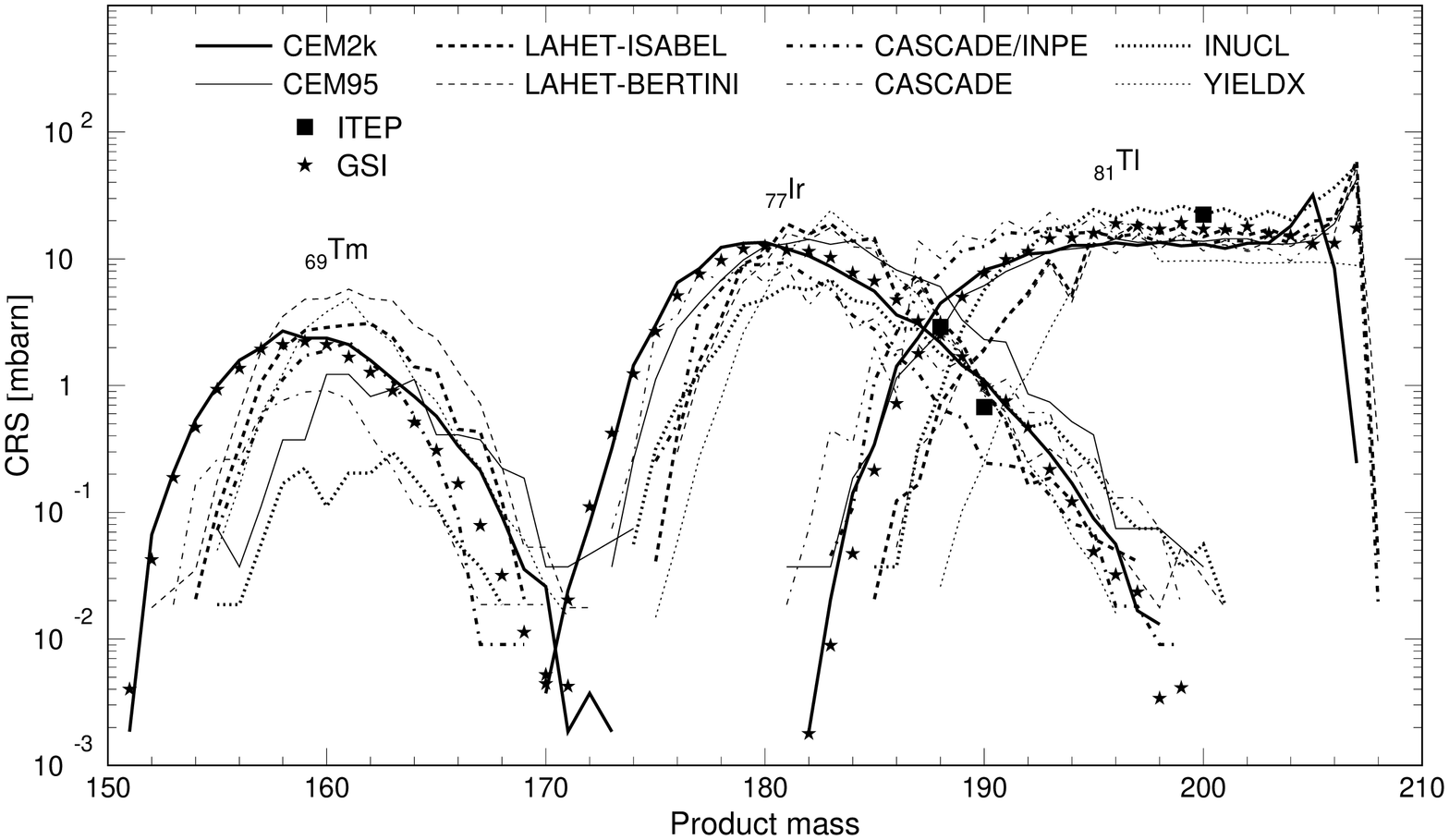}
\vspace*{2mm}
\caption{Isotopic mass distribution for independent products
of Tm, Ir, and Tl isotopes. Black squares are our measurements,
while filled stars show GSI data obtained in reverse kinematics.
Results from different codes are marked as indicated.
}
\label{fig6b}
\end{figure*}
\end{center}

\begin{center}
\begin{figure*} 
\includegraphics[angle=-90,width=17cm]{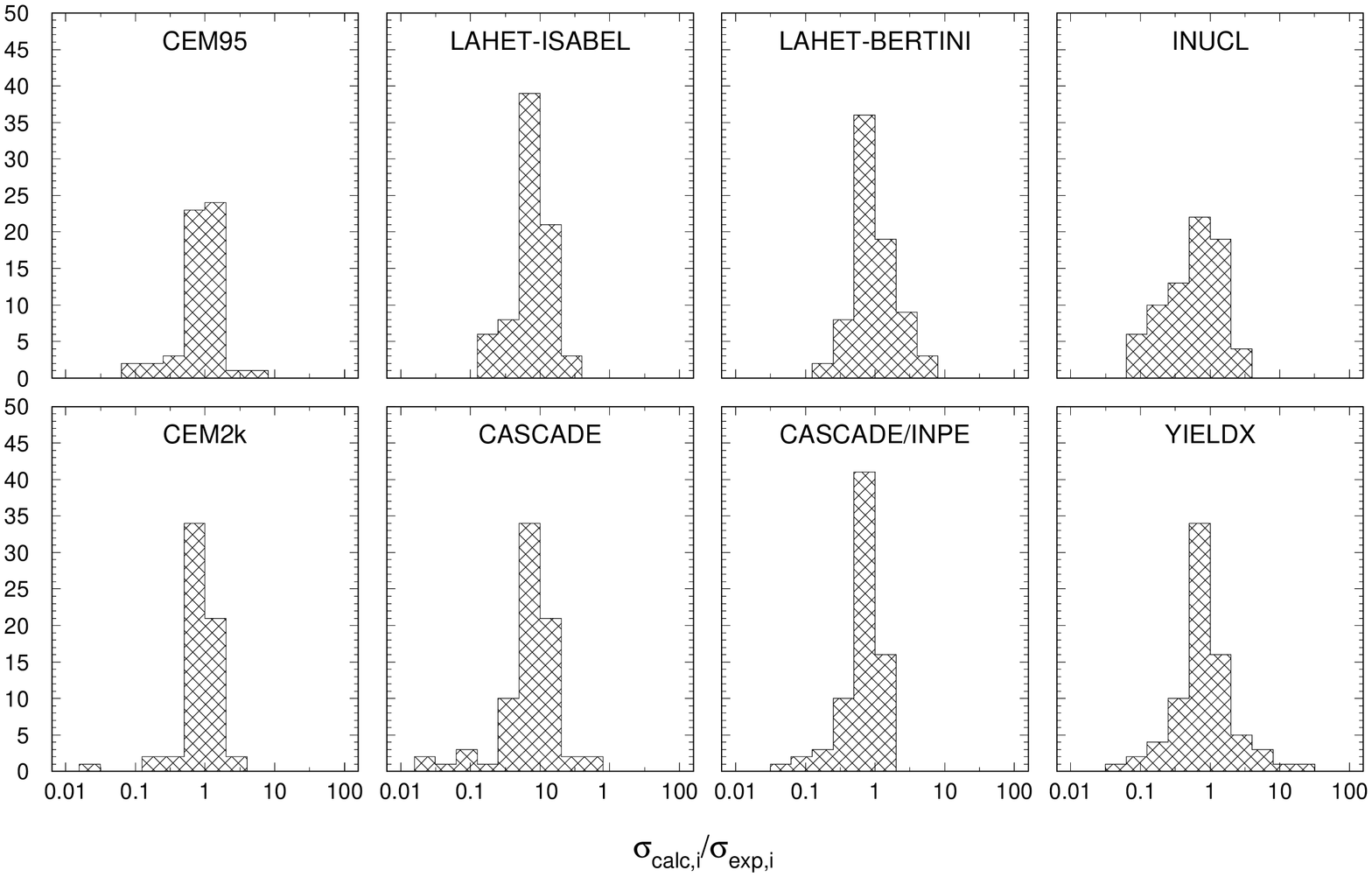}
\caption{Statistics of the simulation-to-experiment ratios (criterion 2).}
\label{fig7}
\includegraphics[angle=-90,width=17cm]{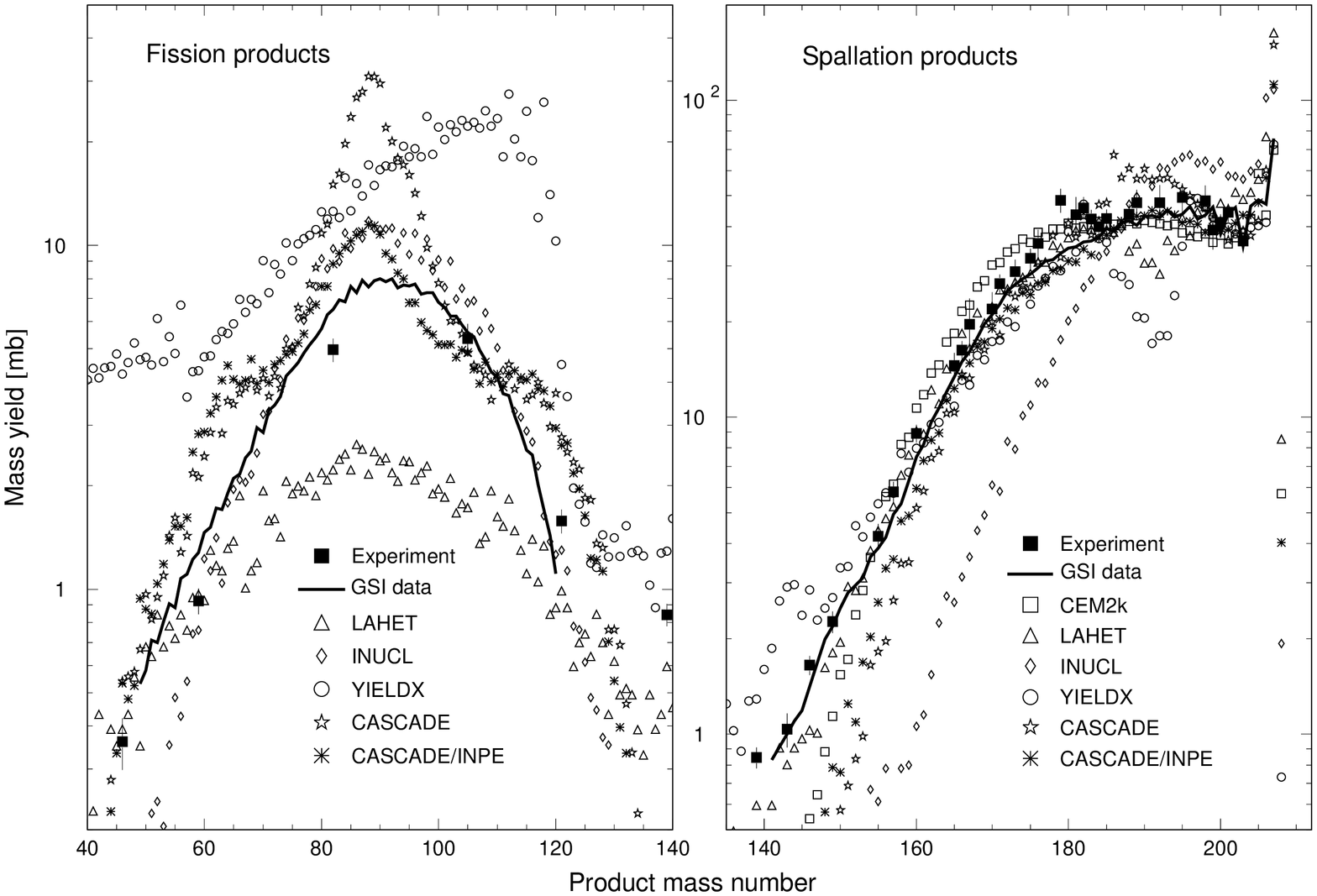}
\vspace*{2mm}
\caption{The simulated mass distributions of reaction products together 
with the measured cumulative and supra-cumulative yields. Black line 
shows GSI data in reverse kinematics.}
\label{fig8}
\end{figure*}
\end{center}

The results obtained with the codes are presented in
\begin{itemize}
\item 
Fig.~\ref{fig6}, which shows the results of a detailed comparison 
between simulated and experimental 
independent and cumulative products (criterion 1);
\item 
Fig.~\ref{fig6b}, which shows the results of a detailed comparison 
between simulated and experimental independent products 
of all isotopes of Tm, Ir, and Tl measured
in this experiment (black squares) together with the data obtained
by the reverse-kinematics method at GSI (black stars) \cite{8}; 
\item 
Fig.~\ref{fig7}, which shows the statistics of the 
simulated-to-experimental data ratios (criterion 2);
\item 
Fig.~\ref{fig8}, which shows the simulated mass distributions of the 
products together with the 
measured cumulative and supra cumulative yields of nuclides 
that are in immediate 
proximity to the stable isotope of a given mass (the sum of 
such yields from either side in     
cases when both left- and right-hand branches of the chain are present).
Obviously, the simulation results do not contradict the 
experimental data if calculated values run 
above the experimental data and follow a general trend of the latter.
This is because direct $\gamma$ spectrometry identifies only the 
radioactive products, which generally form 
a significant fraction of the total mass yield but 
are never equal to the total mass yield when a stable isobar is produced.
\end{itemize}

Table III
presents quantitative information concerning the
agreement of the simulated yields with experimental data for each of
the simulation codes, namely:
\begin{itemize}
\item 
the total number of measured yields, N$_T$;
\item 
of them, the number of the measured yields selected to compare with 
calculations, N$_G$. 
We reject the following
nuclides from our comparison in the cases where: 
\begin{enumerate}
\item 
The measured product is of metastable or just ground state, namely,
\nuc{204m}{Pb}, \nuc{197m}{Pb}, \nuc{198m1}{Tl}, \nuc{196m}{Tl}, \nuc{197m}{Hg}, 
\nuc{195m}{Hg}, \nuc{193m}{Hg}, \nuc{198m}{Au}, \nuc{198g}{Au}, \nuc{186g}{Ir}, 
\nuc{183m}{Os}, \nuc{182m}{Re}, \nuc{121m}{Te}, \nuc{119m}{Te}, \nuc{120m}{Sb}, 
\nuc{114m}{In}, \nuc{110m}{Ag}, \nuc{106m}{Ag}, \nuc{101m}{Rh}, \nuc{90m}{Y}, 
and \nuc{82m}{Rb}; 
\item 
There is a transition of a metastable state to a product out of 
the given decay chain, namely, \nuc{198}{Tl}, \nuc{190}{Ir}, \nuc{152}{Tb}, 
\nuc{149}{Gd},
\nuc{121}{Te}, \nuc{96}{Tc}, \nuc{95}{Tc}, \nuc{95}{Nb}, \nuc{95}{Nb}, 
\nuc{89}{Zr}, \nuc{87}{Y}, and \nuc{85}{Sr};
\item 
There is a strong correlation between a measured cumulative yield 
decaying into another one, namely, 
\nuc{173}{Ta} $\longrightarrow$ \nuc{173}{Hg}, \nuc{170}{Hf} $\longrightarrow$ \nuc{170}{Lu}, 
\nuc{169}{Lu} $\longrightarrow$ \nuc{169}{Yb}, and
\nuc{153}{Tb} $\longrightarrow$ \nuc{153}{Gd}.
The cumulative yields of the precursors in all the above chains  
are almost equal 
to the cumulative yields of the daughters, that is why only the daughter 
yields were used in our comparison,
to prevent double counting. 
Also,  in case of a strong correlation between the cumulative and independent 
yields of a product (\nuc{88}{Y}), only the
independent yield was used for comparison.
\end{enumerate}

\item 
of them, the number of the product nuclei whose yields were simulated 
by a particular code, N$_S$;
\item 
the number of comparison events when the simulated results differ from the 
experimental data by not more than 30\%, N$_{C_{1.3}}$, and the number of  
comparison events when the calculations differ from data 
by not more than a factor of 2.0, N$_{C_{2.0}}$;
\item 
the mean squared deviation of the simulated results from experimental data, 
$<F>$, and its standard deviation, $S(<F>)$.
\end{itemize}

\begin{table} 
\label{code_comp}
\caption{Statistics of comparison between experimental and 
simulated yields in 1.0 GeV proton-irradiated $^{208}${Pb}} 
\vspace{8pt}   
\begin{center}
\begin{tabular}{|c|c|rl|}
&\multicolumn{3}{|c|}{N$_T$ = 114, N$_G$ = 76} \\ \cline{2-4}
Code & N$_{C_{1.3}}$/&$<F>$&$S(<F>)$\\ 
& N$_{C_{2.0}}$/N$_{S}$ & &\\ \hline
LAHET-ISABEL&	39/59/76	&	1.92	&	1.73\\ \hline
LAHET-Bertini & 33/54/76       &        2.04   &        1.71 \\ \hline
CEM95	&	30/46/55	&	2.16	&	2.00\\ \hline
CASCADE&	29/55/72	&	2.18	&	1.88\\ \hline
CASCADE/INPE&   32/56/70	&	1.85	&	1.59\\ \hline
INUCL	&	24/40/73	&	2.89	&	2.15\\ \hline
YIELDX	&	27/49/76	&	2.76	&	2.24\\ \hline
\hline
CEM2k	&	33/54/60	&	1.62	&	1.45\\ 
\end{tabular} 
\end{center}
\end{table}

Since about a third of all secondary nuclei from our reaction
are not spallation products, the ability of codes to   
simulate high-energy fission 
processes is an important criterion for their ability to work
when the target is heavy enough to fission.
Among the codes used here,
LAHET, CASCADE, INUCL, CASCADE/INPE, and YIELDX simulate both 
spallation and fission products.
The CEM95 and CEM2k codes simulate spallation only and
do not calculate the process of fission, and do
not provide fission fragments and a further possible evaporation of
particles from them. When, during a Monte Carlo simulation of the
compound stage of a reaction these codes encounter a
fission, they simply remember this event (that 
permits them to calculate fission cross section and fissility)
and finish the calculation of this event 
without a subsequent calculation of fission fragments.
Therefore, results from CEM95 and CEM2k shown here
reflect the contribution to
the total yields of the nuclides only from deep spallation processes of 
successive emission of particles from the target, 
but do not contain fission products.
This is explicitly reflected in a smaller number of the products 
simulated (the quantity N$_S$ in Table III
and in the 
shapes of the 
simulation curves in Figs.~\ref{fig6} and \ref{fig8}).
To be able to describe 
nuclide production in the fission region, these codes have to be extended
by incorporating a model of high energy fission
(e.g., in the transport code MCNPX \cite{mcnpx}, where CEM97 is used, 
it is supplemented by the RAL fission model \cite{ral}).
 
The following conclusions follow from the analysis of the results 
presented in Table III
and in Figs.~6--9:
\begin{enumerate}
\item 
All codes can reasonably adequately simulate the weak spallation reactions 
(the A$>$180 products), with simulation results differing from 
experimental data, usually within a factor of 2. The exceptions are the 
yields of $^{203}${Hg} (underestimated by an order of magnitude by  
YIELDX) and $^{188}${Ir} (underestimated by an order of magnitude by 
CASCADE/INPE). It should be noted that all the codes simulate a number of 
products ($^{184}${Ir}, $^{198}${Tl}, $^{199}${Tl}, $^{200}${Tl}(cum), 
$^{201}${Tl}, and $^{203}${Pb}), which differ from experiment by less 
than 30\%.
\item 
In the range of the deep spallation reactions (150$<$A$<$180), the 
simulation codes have very different predictive powers, namely,
\begin{itemize}
\item 
the LAHET, CASCADE/INPE, and YIELDX predictions are very close to the 
experimental data; the only exception is $^{172}${Lu}(ind), whose measured 
yield is about a factor of two smaller than LAHET and YIELDX 
give and about a factor of two 
higher than the CASCADE/INPE prediction;
\item  
the CASCADE code simulates the A$>$160 product yields quite adequately, 
except for $^{172}$Lu(ind) whose measured yield is about ten times above the 
simulated value; however, as the atomic number of the product decreases 
below 160, we observe an underestimation of data by CASCADE;
the deviation tends to increase with decreasing A (up to a factor of 5);
\item  
the INUCL code underestimates these reaction products systematically by a 
factor of 2--10 with the discrepancy increasing with decreasing A;
\item
the new code CEM2k was developed taking into account the recent GSI 
measurements \cite{8}. In the spallation region it agrees best with the
data compared to all the other codes, although like its predecessor CEM95,
it does not contain explicit treatment of fission fragments and 
should be supplemented in a transport code by a model of fission-fragment 
formation, or should be improved further by the development
and incorporation of its own  model of fission-fragment formation.
\end{itemize}
\item 
In the range of fission products (40$<$A$<$150), the INUCL code predictions 
are in the best agreement with the data. As a rule, the INUCL-simulated 
yields differ from measured data by a factor of less than 1.5, except for 
$^{59}${Fe}, $^{139}${Ce}, and $^{143}${Pm}. The LAHET-simulated yields 
underestimate data by a factor of 1.5--10, except for $^{88}${Zr} and 
$^{105}${Ag}. The specific predictions of isotope production cross 
sections 
(Fig. 6)
of the semi-phenomenological code 
YIELDX both under- and overestimate the fission product data
by a factor of up to 30, without showing any obvious patterns in the
disagreement. In contrast, the YIELDX isobar cross sections are all greatly
overpredicted, as shown in Fig.~\ref{fig8}. The CASCADE/INPE-simulated
yields of the 130$<$A$<$150 products are strongly underestimated 
(up to 1--2 orders of magnitude), 
while the rest of the simulated fission product yields agree with the
experimental data generally within a factor of 2.
As a rule, the agreement of all codes with the data in the fission product
region is worse than in the spallation region; therefore, development of a 
better model for fission-fragment formation is welcomed for any code.
\end{enumerate}

\section{Conclusion}

The interest shown in both the possible transmutation of nuclear 
wastes and the Spallation Neutron Source (SNS) facilities encourage 
us to anticipate that the accumulation and analysis of nuclear data 
for ADS and SNS applications will have the same growth in academic 
interest and 
practical commitments as was the case for nuclear reactor data during the 
last five decades. Therefore, experimental data on the yields of 
proton-induced reaction products as applied to the ADS and SNS
main target and structure materials are of great interest and importance.
It should be emphasized that the charge distributions in the isobaric 
decay chains are important as well. The information thus obtained 
would make it possible, first, to raise the information content of the 
comparisons between experimental and simulated data and, second, to 
reduce the uncertainties in experimental determination of the 
cumulative yields by establishing unambiguous relations between 
$\sigma^{cum}$ and $\sigma^{cum^{\displaystyle*}}$ for many of the reaction 
products.

We have measured in the present work 114 cross sections of nuclides 
produced in interactions of 1 GeV protons with $^{208}$Pb, of which
8 are independent yields of ground states,
15 are independent yields of metastable states,
15 are independent yields of metastable and ground states,
65 are cumulative yields, and 11 are supra cumulative yields.
We have compared our data with previous measurements and with 
predictions of seven different codes used
in many current applications to understand qualitatively and to
estimate quantitatively their predictive powers.

Regarding the codes benchmarked here, we conclude that none
of them agree well with the data in the whole mass region
of product nuclides and all should be improved further.
In addition, the predictive power of all codes for data in the fission
product
region is worse than in the spallation region; therefore, development of 
better models for fission-fragment formation is of first priority.
The new CEM2k code developed recently at Los Alamos \cite{cem2k}
motivated by the recent GSI data \cite{8} agrees  best with our data
in the spallation region, of the codes tested.
%But CEM2k has yet to be completed by a model of fission-fragment 
%formation, not yet being applicable in the fission-product region.
But CEM2k is inapplicable in the fission-product region,
as to date it has no model of fission-fragment formation.

\section*{Acknowledgments}

The authors are indebted 
to Drs.~T.~Enqvist and B.~Mustapha
for providing us the cross sections measured at the GSI,
to Prof.~R.~Michel for 
sending us the nuclide production data obtained at the ZSR,
to Dr.~F.~E.~Chukreev for helpful 
comments on the nuclear decay-chain data,
and to Prof.~V.~Artisyuk for useful discussions and help.

The work has been performed under the ISTC Project \#839 supported by
the European Community, Japan (JAERI), and Norway and was partially
supported by the U.~S.~Department of Energy.

\end{document}